\documentclass[iop,apj,numberedappendix,appendixfloats]{emulateapj}

\usepackage{natbib}
\usepackage{longtable}
\usepackage{rotating}
\usepackage[usenames,dvipsnames]{color}
\usepackage{enumerate}
\usepackage{amsmath}
\usepackage{multirow}

\usepackage[colorlinks=true,citecolor=blue,linkcolor=blue]{hyperref} 

\usepackage{threeparttable}
\usepackage{txfonts}
\usepackage{array} 
\usepackage{CJK}

\def\g0{G$_{0}$}
\def\2cm{cm$^{-2}$}
\def\cm3{cm$^{-3}$}

\def\nh3{NH$_3$}
\def\n2h{N$_2$H$^+$}
\def\C2h{C$_2$H}

\def\13co{$^{13}$CO}
\def\c18o{C$^{18}$O}
\def\hc3n{HC$_3$N}
\def\h2{H$_2$}
\def\nh{n(H$_2$)}

\def\ch3p{CH$_{3}^{+}$}

\shorttitle{}
\shortauthors{}
\slugcomment{}

\begin{document}

\title{PSR J1926$-$0652: a pulsar with interesting emission properties discovered at FAST}
\author{Lei Zhang \altaffilmark{1,2,3}{*},
Di Li \altaffilmark{1,2,4}{*}, 
George Hobbs \altaffilmark{3,1}{*}, 
Crispin H. Agar\altaffilmark{5},
Richard N. Manchester\altaffilmark{3},
Patrick Weltevrede\altaffilmark{5},
William A. Coles \altaffilmark{6},
Pei Wang \altaffilmark{1}, 
Weiwei Zhu \altaffilmark{1}, 
Zhigang Wen \altaffilmark{7}, 
Jianping Yuan  \altaffilmark{7}, 
Andrew D. Cameron\altaffilmark{3,1},
Shi Dai \altaffilmark{3,1},
Kuo Liu \altaffilmark{8,1}, 
Qijun Zhi \altaffilmark{9,10}
Chenchen Miao \altaffilmark{1,2}
Mao Yuan \altaffilmark{1,2}
Shuyun Cao\altaffilmark{4},
Li Feng\altaffilmark{4},
Hengqian Gan\altaffilmark{4},
Long Gao\altaffilmark{4},
Xuedong Gu\altaffilmark{4},
Minglei Guo\altaffilmark{4},
Qiaoli Hao\altaffilmark{4},
Lin Huang\altaffilmark{4},
Peng Jiang\altaffilmark{4},
Chengjin Jin\altaffilmark{4},
Hui Li\altaffilmark{4},
Qi Li\altaffilmark{4},
Qisheng Li\altaffilmark{4},
Hongfei Liu\altaffilmark{4},
Gaofeng Pan\altaffilmark{4},
Zhichen Pan\altaffilmark{4},
Bo Peng\altaffilmark{4},
Hui Qian\altaffilmark{4},
Lei Qian\altaffilmark{4},
Xiangwei Shi\altaffilmark{4}, 
Jinyou Song \altaffilmark{4},
Liqiang Song\altaffilmark{4},
Caihong Sun\altaffilmark{4},
Jinghai Sun\altaffilmark{4},
Hong Wang\altaffilmark{4},
Qiming Wang\altaffilmark{4},
Yi Wang\altaffilmark{4},
Xiaoyao Xie\altaffilmark{4},
Jun Yan\altaffilmark{4},
Li Yang\altaffilmark{4},
Shimo Yang\altaffilmark{4},
Rui Yao\altaffilmark{4},
Dongjun Yu\altaffilmark{4},
Jinglong  Yu\altaffilmark{4},
Youling Yue\altaffilmark{4},
Chengmin Zhang\altaffilmark{4},
Haiyan Zhang\altaffilmark{4},
Shuxin Zhang\altaffilmark{4},
Xiaonian Zheng\altaffilmark{4},
Aiying Zhou\altaffilmark{4},
Boqin Zhu\altaffilmark{4},
Lichun Zhu\altaffilmark{4},
Ming Zhu\altaffilmark{4},
Wenbai Zhu\altaffilmark{4},
Yan Zhu \altaffilmark{4}
}
\affil{
$^1$ National Astronomical Observatories, Chinese Academy of Sciences, A20 Datun Road, Chaoyang District, Beijing 100101, China\\
$^2$ University of Chinese Academy of Sciences, Beijing 100049, China\\
$^3$ CSIRO Astronomy and Space Science, PO Box 76, Epping, NSW 1710, Australia\\
$^4$ FAST Collabration, CAS Key Laboratory of FAST, NAOC, Chinese Academy of Sciences, Beijing 100101, China\\
$^5$ Jodrell Bank Centre for Astrophysics, The University of Manchester, UK\\
$^6$ ECE Department, University of California at San Diego, La Jolla, CA, USA\\
$^7$ Xinjiang Astronomical Observatory, 150, Science-1 Street, Urumqi, 830011 Xinjiang,  China\\
$^8$ Max-Planck-Institut f\"{u}r Radioastronomie, Auf dem H\"{u}gel 69, D-53121 Bonn, Germany\\
$^9$ Guizhou Provincial Key Laboratory of Radio Astronomy and Data Processing, Guizhou Normal University, Guiyang, China, 550001\\
$^{10}$ School of Physics and Electronic Science, Guizhou Normal University, Guiyang, China, 550001
}
\email{{*}Email: leizhang996@nao.cas.cn, dili@nao.cas.cn,George.Hobbs@csiro.au}

\begin{abstract}
We describe PSR~J1926$-$0652, a pulsar recently discovered with the Five-hundred-meter Aperture Spherical radio Telescope (FAST). Using sensitive single-pulse detections from FAST and long-term timing observations from the Parkes 64-m radio telescope, we probed phenomena on both long and short time scales. The FAST observations covered a wide frequency range from 270 to 800\,MHz, enabling individual pulses to be studied in detail. The pulsar exhibits at least four profile components, short-term nulling lasting from 4 to 450 pulses, complex subpulse drifting behaviours and intermittency on scales of tens of minutes.  While the average band spacing $P_3$ is relatively constant across different bursts and components, significant variations in the separation of adjacent bands are seen, especially near the beginning and end of a burst. Band shapes and slopes are quite variable, especially for the trailing components and for the shorter bursts. We show that for each burst the last detectable pulse prior to emission ceasing has different properties compared to other pulses. These complexities pose challenges for the classic carousel-type models.
\end{abstract}


\section{Introduction}

The Five-hundred-meter Aperture Spherical radio Telescope (FAST), located in southern China, is the world's largest single-dish radio telescope. Between August 2017 and February 2018, FAST carried out drift-scan observations using an ultra-wide-bandwidth receiver (UWB) to search for radio pulsars. Before the removal of this UWB in May 2018 (it was installed primarily for commissioning projects), 60 pulsar candidates had been obtained, with 44 of these already confirmed either by further FAST observations or using the Parkes 64-m or Effelsberg 100-m radio telescopes. The survey strategy and the full collection of new discoveries will be published elsewhere, with a brief summary currently available on the Commensal Radio Astronomy FAST survey (CRAFTS) website\footnote{http://crafts.bao.ac.cn/pulsar/} and in \citet{Qian+19}.

In this paper, we report the discovery of PSR~J1926$-$0652, which has a $\sim1.6$\,s pulse period.  The pulsar was discovered using a single-pulse search pipeline \citep{Zhu+14} that was applied to observations taken in August, 2017.  The pulsar was independently confirmed using the Parkes radio telescope in October, 2017. We have continued observations with both FAST and Parkes and, as described in this paper, show that  PSR~J1926$-$0652 exhibits a wide-range of emission phenomena.

Most pulsars are been discovered through the detection of their regular pulsed signal.  However, soon after the discovery of pulsars it was found that the pulsed emission is seldom completely stable. For example, individual pulses may be significantly stronger than their mean. This led to the development of searches for pulsars through the detection of individual bursts (e.g., \citealp{McLaughlin+06}) and to the single-pulse search pipelines as used for our discovery. 

With sufficiently sensitive telescopes, many pulsars can be shown to have mean pulse profiles formed from subpulses that drift in pulse phase in successive pulses. This phenomenon is known as subpulse drifting and was first reported by \citet{Drake+68}. The subpulse drifting phenomena is often described with a ``carousel model" in which a ring of source regions systematically rotates about the magnetic axis \citep{Ruderman+75}.  We now know that at least one-third of pulsars exhibit subpulse drifting \citep{Weltevrede+06a}. Various algorithms have been developed to quantify the subpulse drifting. For instance, \citet{Edwards+03} show how a two-dimensional fluctuation spectrum (2DFS) can be used to determine the period of the subpulses in both pulse phase (known as $P_2$) and as pulse number ($P_3$, measured in time). They noted that observational results indicated that $P_2$ does not change as a function of observing epoch and $P_3$ does not change with pulse longitude. The drift rate or slope of a subpulse band, is conventionally defined as $\Delta \phi = P_{2}/P_{3}$.

Several pulsars that exhibit complex patterns in their drifting subbands are now known. For instance, \citet{Qiao+04} describe a complex model for the ``bi-drifting" phenomenon seen in PSR~J0815+09.  The observing-frequency-dependence of subpulse drifting has also been studied. For example, \citet{Taylor+75} and \citet{Wolszczan+81} showed that, for PSR B0031$-$07 and PSR B0809+74, $P_2$ varied with frequency approximately as $\nu^{-0.25}$, similar to the dependence expected for radius-to-frequency mapping in polar-cap models of the pulsar emission \citep{Cordes+78}.   

The pulsed emission from some pulsars has also been observed to switch off suddenly. This phenomena, known as ``nulling" was first reported by \citet{Backer+70b}. Nulling is relatively common, particularly in long period pulsars \citep{Rankin+86}. For instance, 43 out of 72 well-observed pulsars were found by \citet{Biggs+92} to exhibit evidence for nulling.  The duration of null events varies widely.  In some cases, one or a few pulses may be missing, whereas in other cases the emission may be undetectable for hours, days, or in extreme cases, months and years.  The ``null fraction" (NF), the fraction of time that the pulsar is in a null state, can range from close to zero (e.g.\ PSR~B1737$+$13; \citealp{Biggs+92}) to more than 90$\%$ \citep{Wang+07}. Pulsars that switch off for very long periods (on scales of hours to years) are often termed ``intermittent pulsars''. \citet{Kramer+06} studied one such pulsar, PSR~B1931$+$24, and showed that the pulsar's slow-down rate is reduced when the pulsar is in its null state. This was explained as a change in magnetospheric currents. Some pulsars are also known to switch between multiple discrete profile states.  This is known as ``mode changing". \citet{Wang+07} and \citet{Lyne+10} suggested that mode changing and nulling are related phenomena and differ only in the magnitude of the changes in the magnetospheric current flows. 

A few papers have described studies into how the nulling and drifting phenomena may be linked. For instance, \citet{Gajjar+17} studied PSRs~J1741$-$0840 and J1840$-$0840. They reported that for PSR~J1840$-$0840 the pulsar tended (but not always) to start nulling after the end of a driftband. When PSR~J1840$-$0840 then switched back on, it typically started at the beginning of a new driftband in both of its profile components. 

Long-term monitoring of a pulsar provides information on the spin-down of the pulsar and long time-scale intermittent behavior. Such monitoring also provides high signal-to-noise, polarisation-calibrated, average pulse profiles that can be used to determine the emission geometry of the system.  The single pulse observations provide information on the nulling and drifting phenomena. Together, these results (for instance, as in \citealt{Rankin+08}) can be used to search for the elusive physical model that will link these emission phenomena.

The pulsar that we describe in this paper, PSR~J1926$-$0652, has multiple pulse profile components and exhibits both subpulse drifting and nulling on various time-scales. The paper is organized as follows.  In Section 2, we describe our observations of PSR~J1926$-$0652. In Section 3, we present our analysis of the individual pulses. In Section 4, we describe the long-term behaviour, the timing solution and analyse the polarization and flux-density properties of the average pulse profile.  We discuss our results in Section 5 and conclude the paper in Section 6.

\section{Observations}

We have carried out observations of PSR~J1926$-$0652 with both the FAST and Parkes radio telescopes.  FAST, which is still being commissioned, has a large collecting area allowing us to observe single pulses from PSR~J1926$-$0652. The Parkes telescope is not sensitive enough to detect single pulses from this pulsar, but can be accurately calibrated and has been used to measure the polarization properties of the pulsar, as well as to carry out long-term monitoring.

\subsection{Observation of single pulses}  

We observed PSR~J1926$-$0652 for $\sim$50\,minutes using FAST on November 28th, 2017 (corresponding to a MJD 58085) using a wide-bandwidth receiver covering from 270\,MHz to 1.6\,GHz.  For most of the early FAST commissioning data, including the observation presented here, only one of the two linear polarization signal paths was reliable and the pulsar was only detectable in the low-frequency band\footnote{It is currently unclear whether this was because the pointing position was inaccurate or whether the telescope efficiency was low in the higher band during these observations.}. We therefore only make use of the low frequency (270-800\,MHz) band with the single available polarization channel. The lack of complete polarization  information limits the scope of our single-pulse analysis. We have therefore focused our analysis on the detectable variations in flux density and in pulse phase. During the observation we recorded a total of 1921 single pulses with a time resolution of 100\,$\mu$s. We subsequently extracted individual pulses\footnote{Note that this requires the use of the -K option.} with 512 phase bins per pulse period using the \textsc{dspsr} program \citep{Straten+11}.

\subsection{Monitoring Observations}

The Parkes telescope continues to be used for regular timing observations of PSR~J1926$-$0652 in the 20\,cm (1400~MHz) observing band with the central beam of the 13-beam receiver \citep{SS+96}. We have obtained 35 observations of this pulsar between 2017 October 8 (MJD 58034) and 2018 September 26 (MJD 58387). Integration times are typically 1\,hour and the observations are divided into 30\,s time segments (known as ``sub-integrations"). The bandwidth used was 256\,MHz, which was divided into 1024 frequency channels and 1024 phase bins were formed across the profile using the Parkes Digital Filterbank Mark 4 (PDFB4). In order to obtain high-quality flux density and polarization calibration solutions, each observation was preceded by observation of a switched calibration noise source. 

We processed the data using the \textsc{psrchive} software suite  \citep{Hotan+04}. Aliased signals and narrow-band radio frequency interference (RFI) were removed by giving zero weight to channels within 5 percent of the band edge and those with a level substantially above a median-smoothed bandpass. PSR~J1926$-$0652 was clearly detected in 29 observations and was undetected on the other six occasions. Since there is no perceptible worsening of RFI conditions in those six epochs, the non-detection is probably due to nulling. The length of observing time for the six non-detections were 11.5\,minutes, 9\,minutes, 17.5\,minutes, 64\,minutes, 55.5\,minutes and 72\,minutes. To convert the measured intensity from the Parkes telescope observations to absolute flux density, we made use of observations of the radio galaxy 3C 218 (Hydra A; \citealp{Baars+77}, see also \citealp{Xie+19})  that are taken every few weeks to support the Parkes Pulsar Timing Array project \citep{Manchester+13}. This allowed us to determine the effective flux density of the calibration noise source and consequently an absolute flux scale.

These long-term observations allowed us to model the rotation of the pulsar (see Section 4.1), to produce a high signal-to-noise ratio (S/N) average pulse profile enabling determination of the polarization properties and the flux density of the pulsar (Section 4.2) and determine the long-term on-off time-scale for the pulse emission (Section 4.3).

The raw and processed data sets described in this paper are available online. See Appendix~\ref{sec:data_access} for details.

\section{The Single-pulse Emission}\label{sec:singlePulse}

\begin{figure}[thp]
\centering
\includegraphics[height=2.5\linewidth,width=1.15\linewidth]{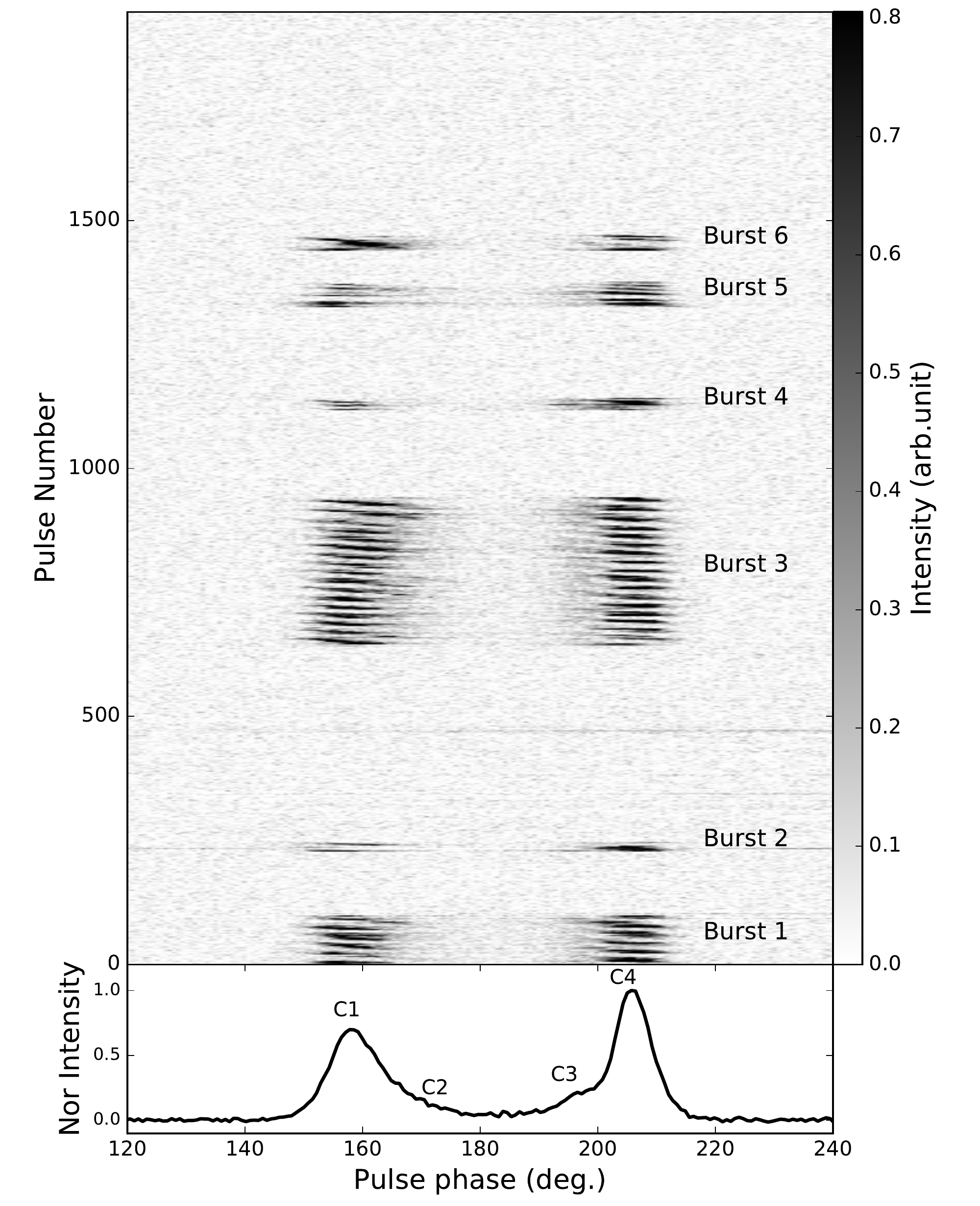}
\caption{(Upper panel) Pulse stack with the single uncalibrated polarization channel averaged across the FAST observing band from 270 to 800\,MHz. The six active intervals are indicated as bursts. (Lower panel) The average pulse profile obtained from these pulses with the four pulse components indicated.}
\label{fig.C12_SP+profile}
\end{figure}

\begin{figure*}[thp]
\centering
\begin{tabular}{ccc}
\includegraphics[height=0.5\linewidth,width=0.33\linewidth]{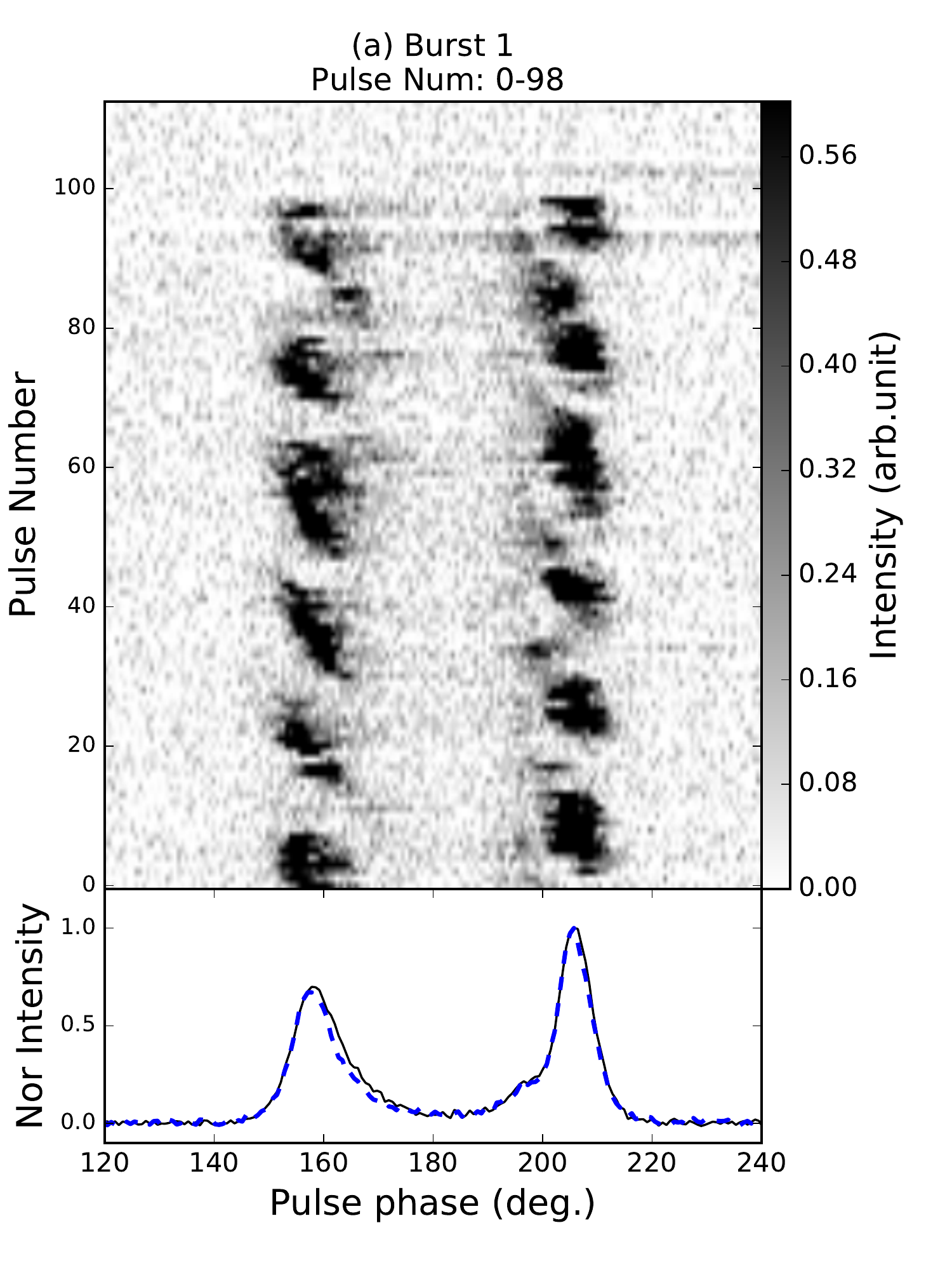}~
\includegraphics[height=0.5\linewidth,width=0.33\linewidth]{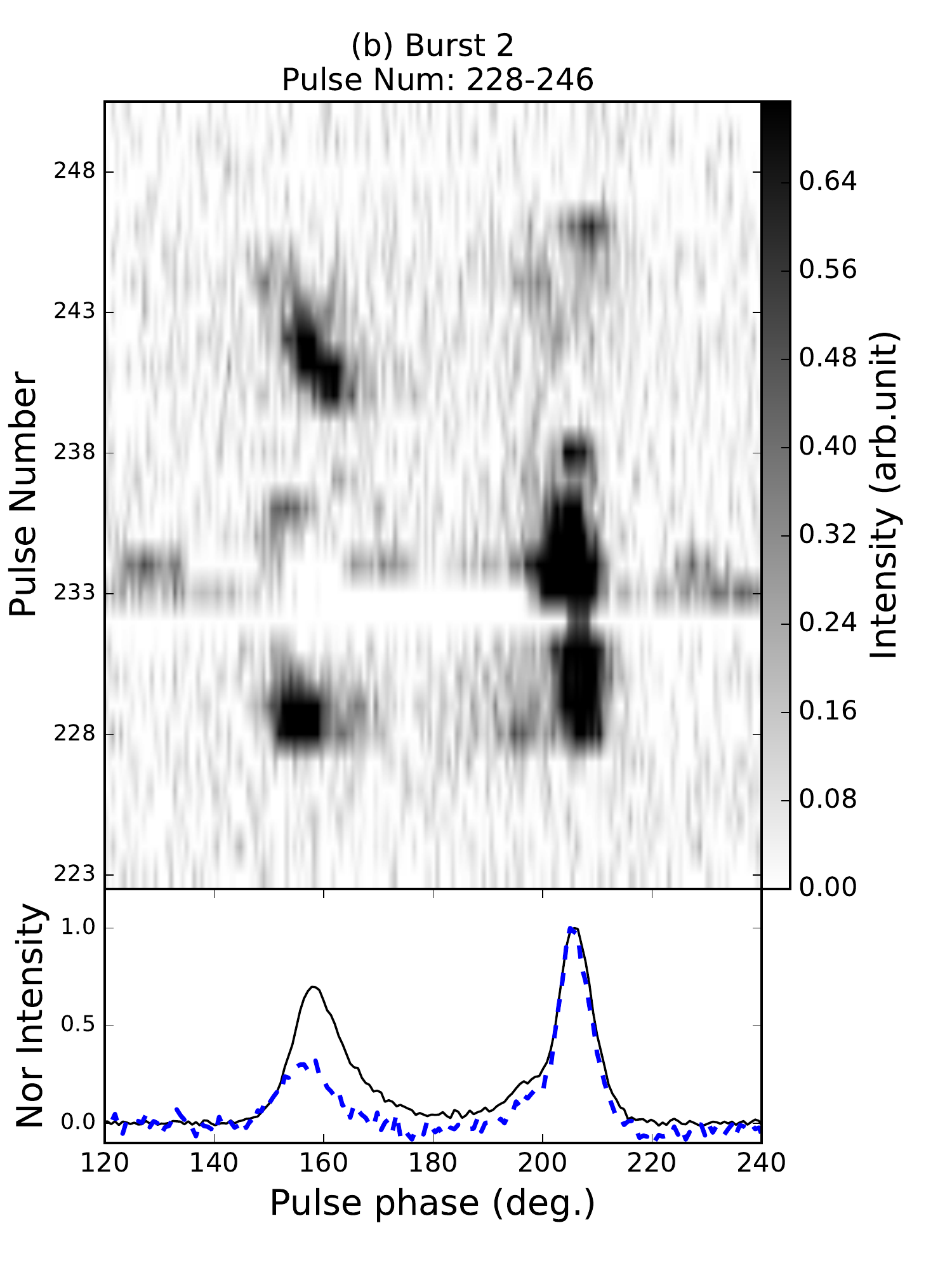}~
\includegraphics[height=0.5\linewidth,width=0.33\linewidth]{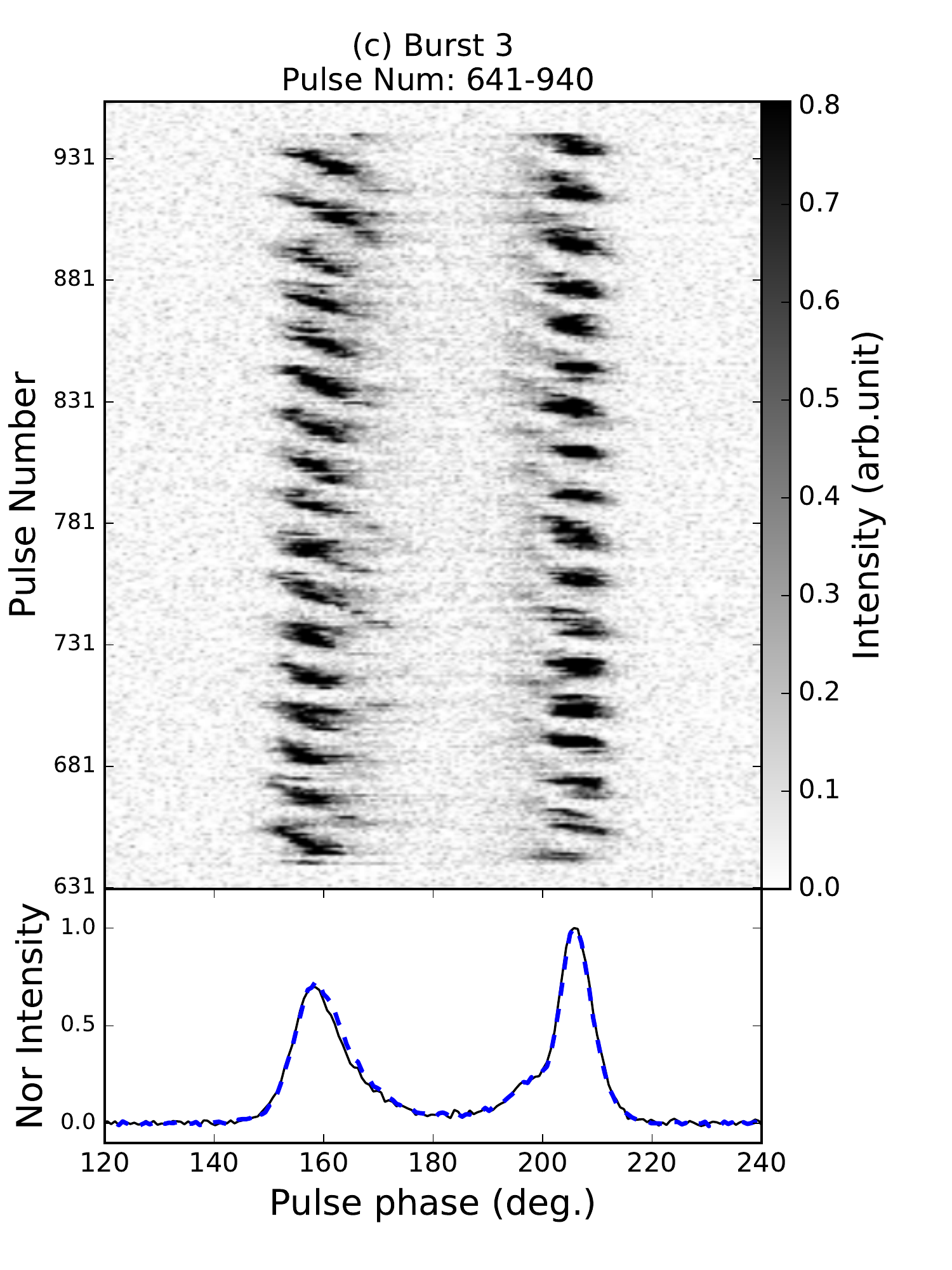}\\
\includegraphics[height=0.5\linewidth,width=0.33\linewidth]{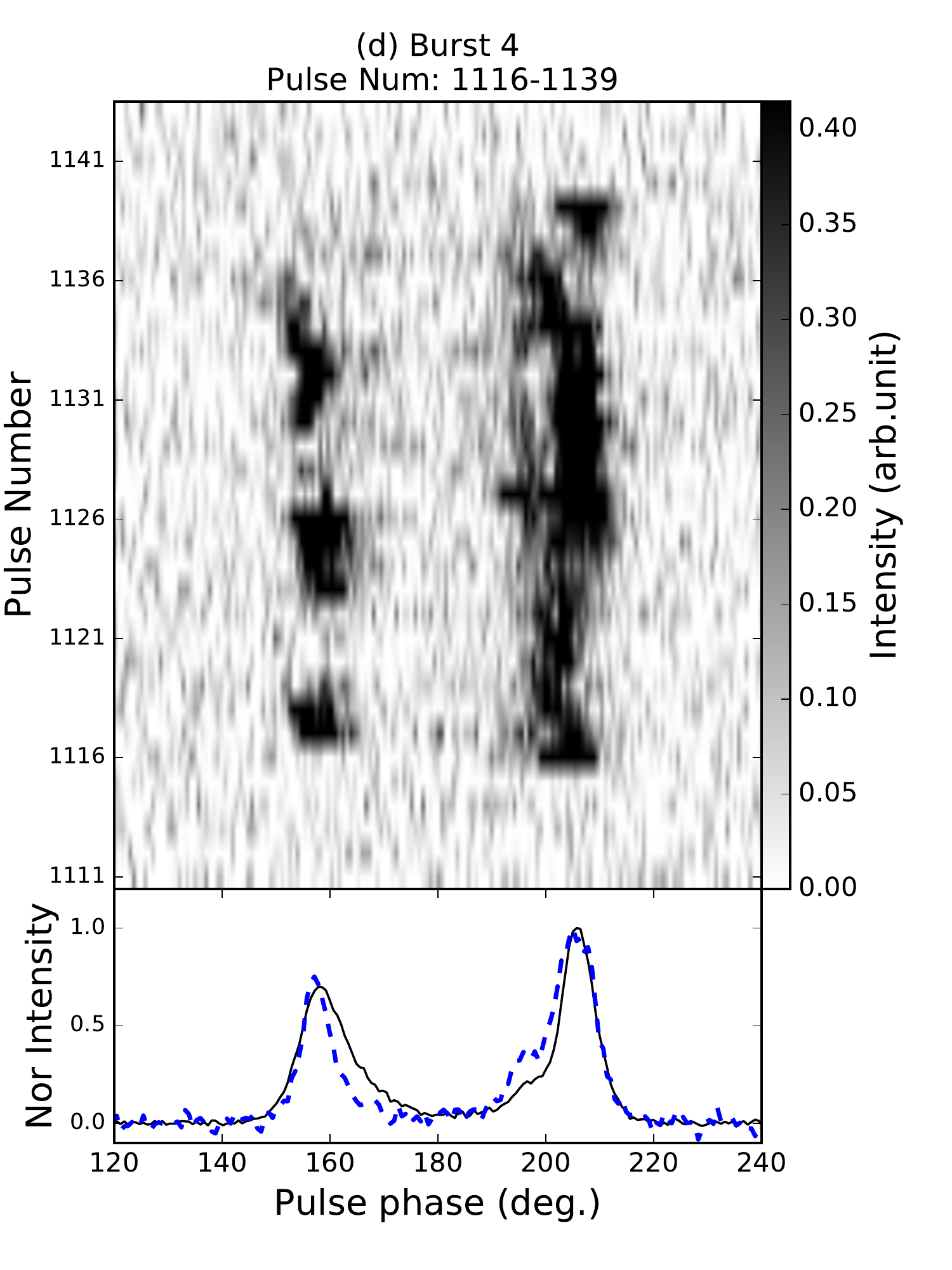}~
\includegraphics[height=0.5\linewidth,width=0.33\linewidth]{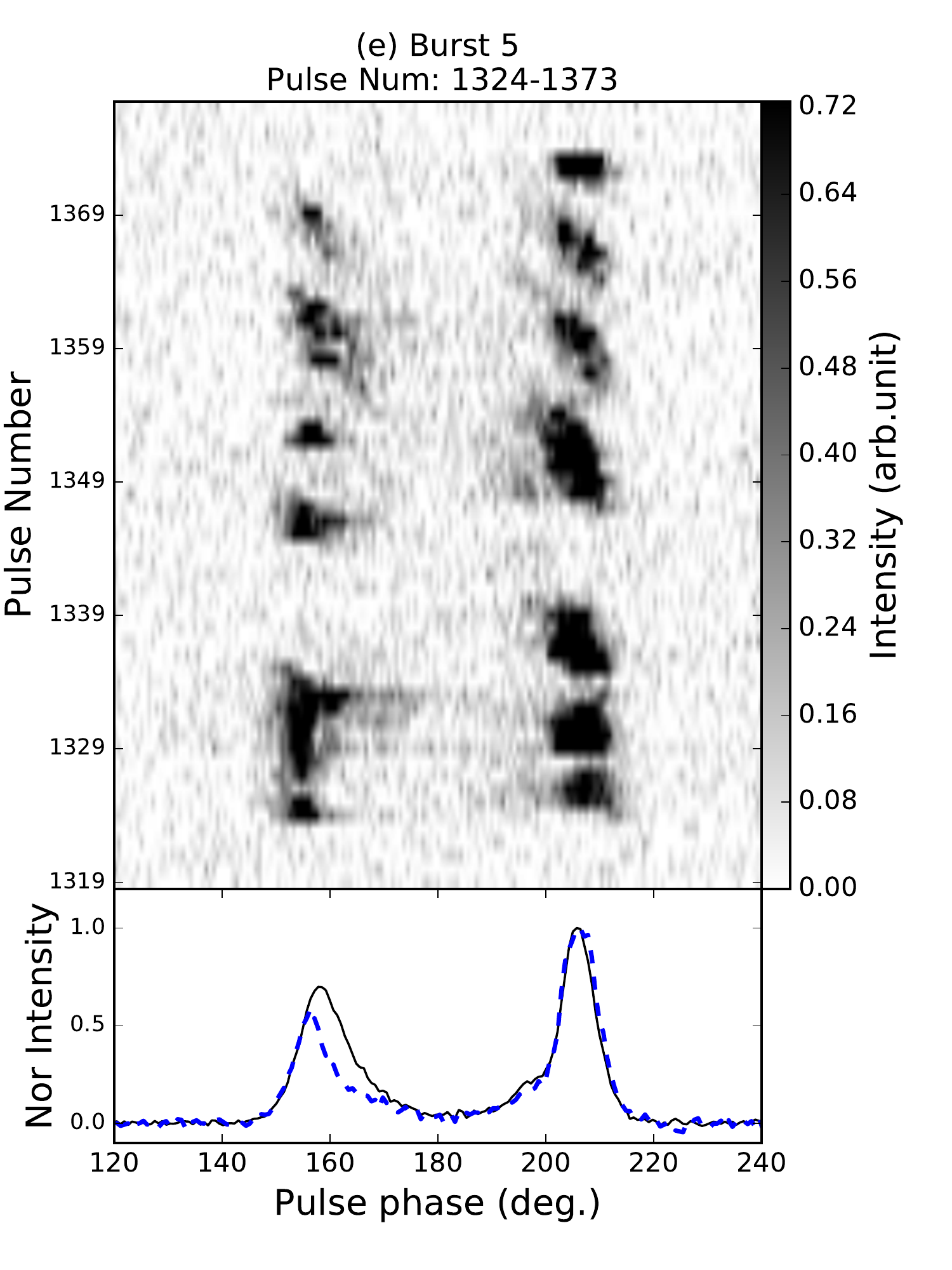}~
\includegraphics[height=0.5\linewidth,width=0.33\linewidth]{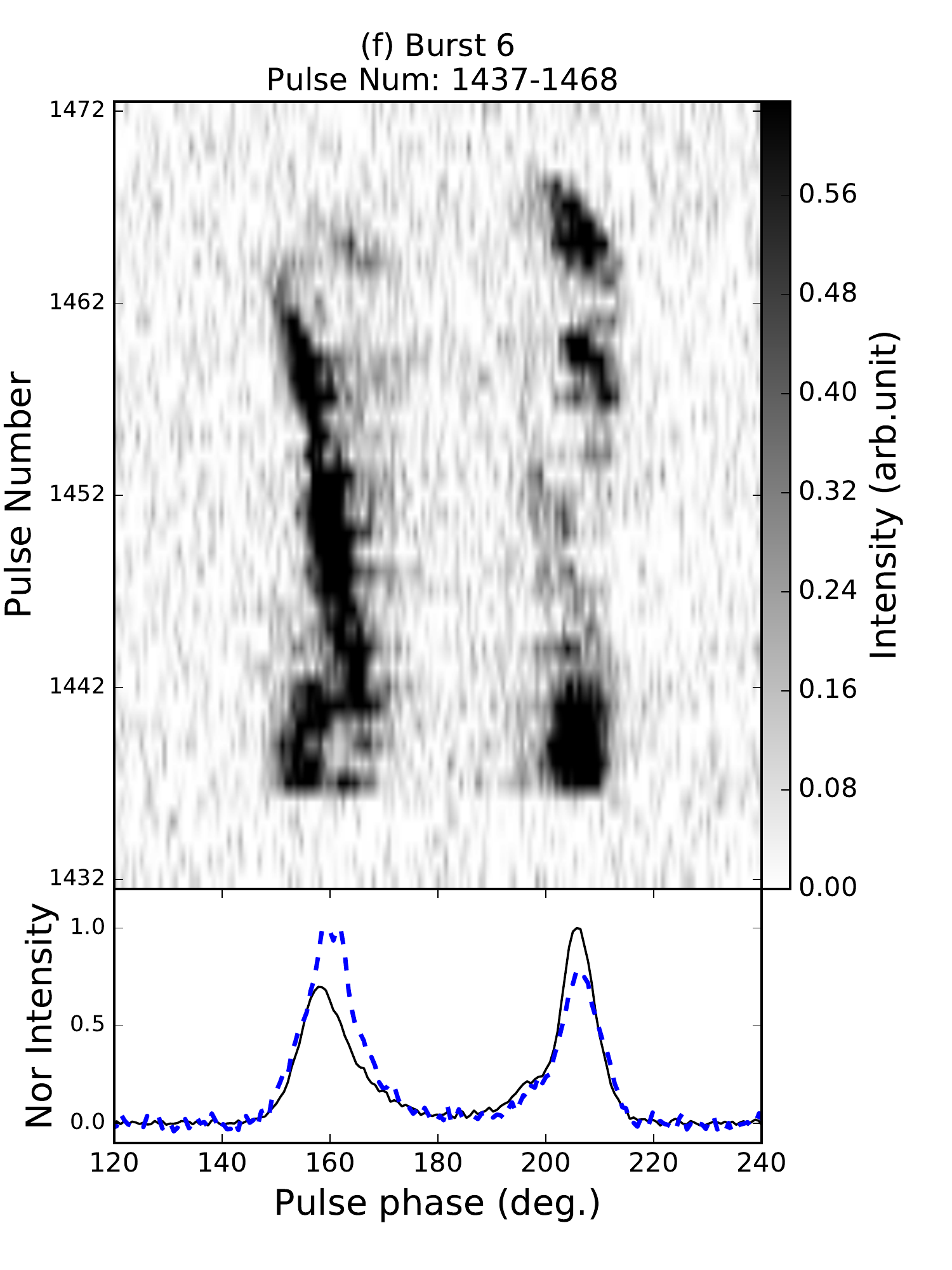}\\
\end{tabular}
\caption{Pulse stacks for each burst are given in the upper panel of each subplot. The lower panel of each subplot shows the pulse profile averaged over the whole data span as a black solid line and averaged over the particular burst state as a blue dashed line.}
\label{fig.C12_6Burst}
\end{figure*}

A ``pulse stack'' is an array of consecutive pulses with pulse phase on the $x$ axis and increasing pulse number on the $y$ axis. The upper panel in Figure~\ref{fig.C12_SP+profile} shows the entire pulse stack obtained using the FAST single-pulse data set across the frequency band from 270 to 800\,MHz with the single available polarization channel.\footnote{In order to show the successive single pulses, we have not removed pulses affected by RFI in the pulse-stack figures (Figure~\ref{fig.C12_SP+profile} and Figure~\ref{fig.C12_6Burst}). However, for the average pulse profiles and the single-pulse analysis described in this paper, we do remove the interference.} The average profile is shown in the lower panel. This profile and pulse stack has been obtained after summing in frequency over the entire band (from 270 to 800\,MHz). 

We have labelled the regions in which the emission is ``on"  in the pulse stack as Bursts 1 to 6.  We show these on-states in more detail in the six panels of  Figure~\ref{fig.C12_6Burst}. The average pulse profile from each of these bursts is shown in the lower section of each panel overlaid on the mean pulse profile for the whole observation. Various emission phenomena are seen in these panels including multiple profile components, subpulse drifting and nulling. 

The average pulse profile consists of two main components (labelled as C1 and C4 in the lower panel of Figure~\ref{fig.C12_SP+profile}). An inspection of Figure~\ref{fig.C12_6Burst} shows at least two extra components.  A weak component to the right of C1 is seen in several bursts (we label this component C2) and similarly, a weak component to the left of C4 leads to the ``bump" in the average profile that we have labelled C3. See Appendix~\ref{sec:P3fold} for details.  Between these components there is a bridge region of emission.

\begin{figure}[thp]
\centering
\includegraphics[width=1.0\linewidth]{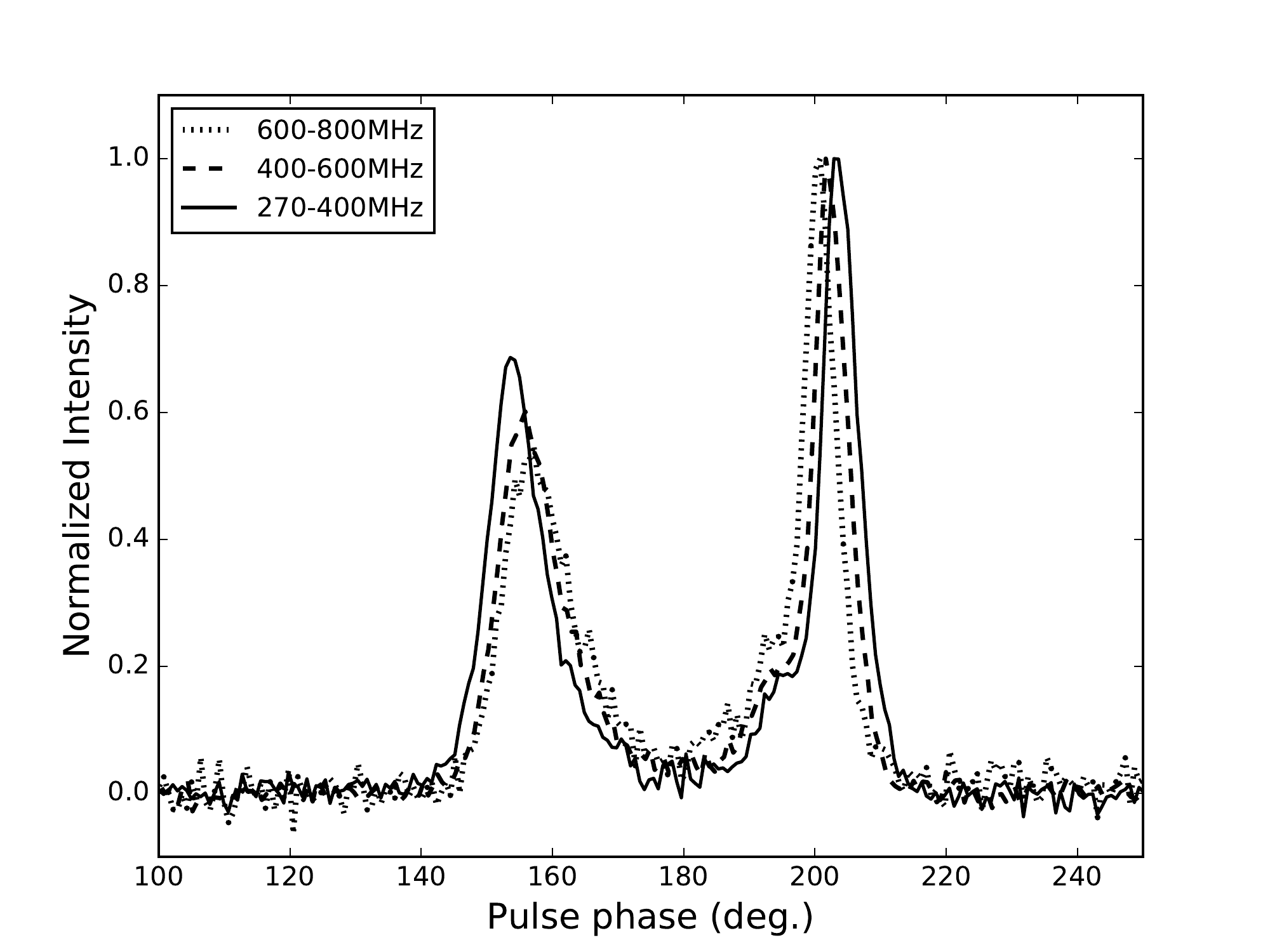}
\caption{Mean pulse profiles for the whole observation in three frequency subbands across the observed bandwidth of 270 to 800~MHz. The profiles are normalised to a peak amplitude of 1.0 and  have been aligned by the midpoint of the leading and trailing edges at 50\% of the profile peak.}
\label{C12_MFprof}
\end{figure}

Over the wide observed FAST band (270 to 800\,MHz) we expect to see pulse shape evolution relating to intrinsic profile changes, emission arising from different positions in the magnetosphere and interstellar-medium effects. Figure~\ref{C12_MFprof} shows mean pulse profiles for three subbands across this observed bandwidth. It is clear that, as the observing frequency increases, the component separation decreases -- pulse widths at 50\% of the peak amplitude for the low and high frequency bands are given in Table~\ref{tab.C12_parameters}. The reduction in profile width as a function of frequency is well-known in the general pulsar population and is usually attributed to radius-to-frequency mapping \citep{Cordes+78}.

As Figures~\ref{fig.C12_SP+profile} and \ref{fig.C12_6Burst} show, the observed burst durations cover a wide range. It could be argued that Burst 5 is in fact two bursts, 5a and 5b, separated by a null of four pulse periods. Given this, the burst durations range from 17 pulse periods for Burst 5a to 300 pulse periods for Burst 3.\footnote{If Burst 5 is treated as a single burst, the minimum burst duration is 29 pulse periods for Burst 2.} The null durations are also highly variable and range from four to more than 450 pulse periods. The pulsar is in a null state about 75\% of the time, but this null fraction is quite uncertain because of the limited number of bursts observed.

One striking property of the emission is that during the longest burst event, Burst 3, the leading components drift later in phase, with the separation of C1/C2 from C3/C4 decreasing through the burst. This is most easily seen in Figure~\ref{fig.C12_SP+profile}. Similar behaviour may be occurring in other bursts, but this is not certain. It appears that the phase of components C1/C2 resets to the same starting value for each burst.  Longer data sets are needed to confirm this property and to investigate it in more detail.

\begin{figure}
\centering
\includegraphics[width=1\linewidth]{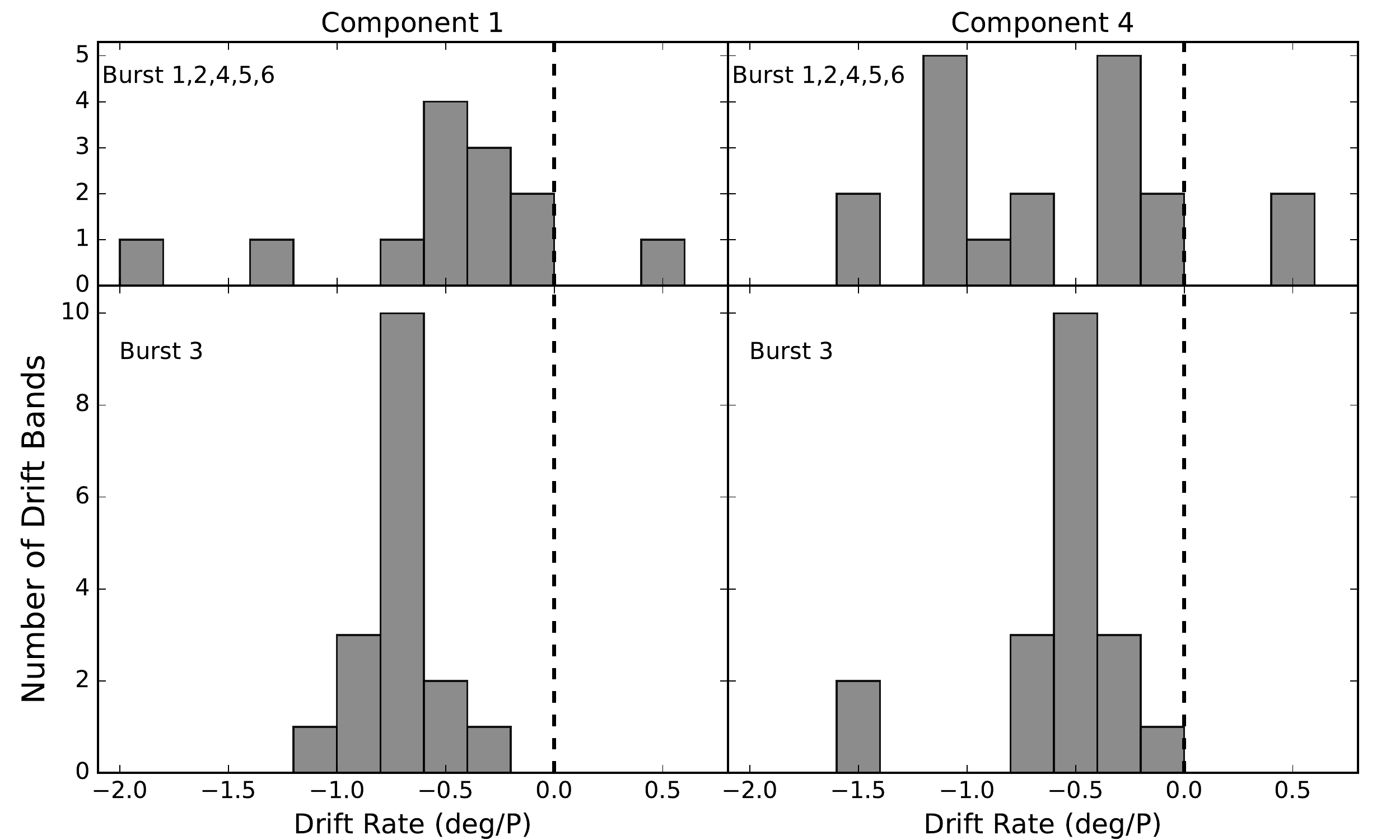}
\caption{Histograms of the observed subpulse drift rates or band slopes for all observed drift bands, separately for components C1 and C4. The bin width is the same for all histograms ($0.2\degr /P$) and has been chosen to approximate the typical uncertainty in the measured band slopes. The vertical dashed lines mark zero drift rate. }
\label{fig.BursHis}
\end{figure}

\begin{figure*}[thpb]
\centering
\begin{tabular}{ll}
\includegraphics[width=8cm]{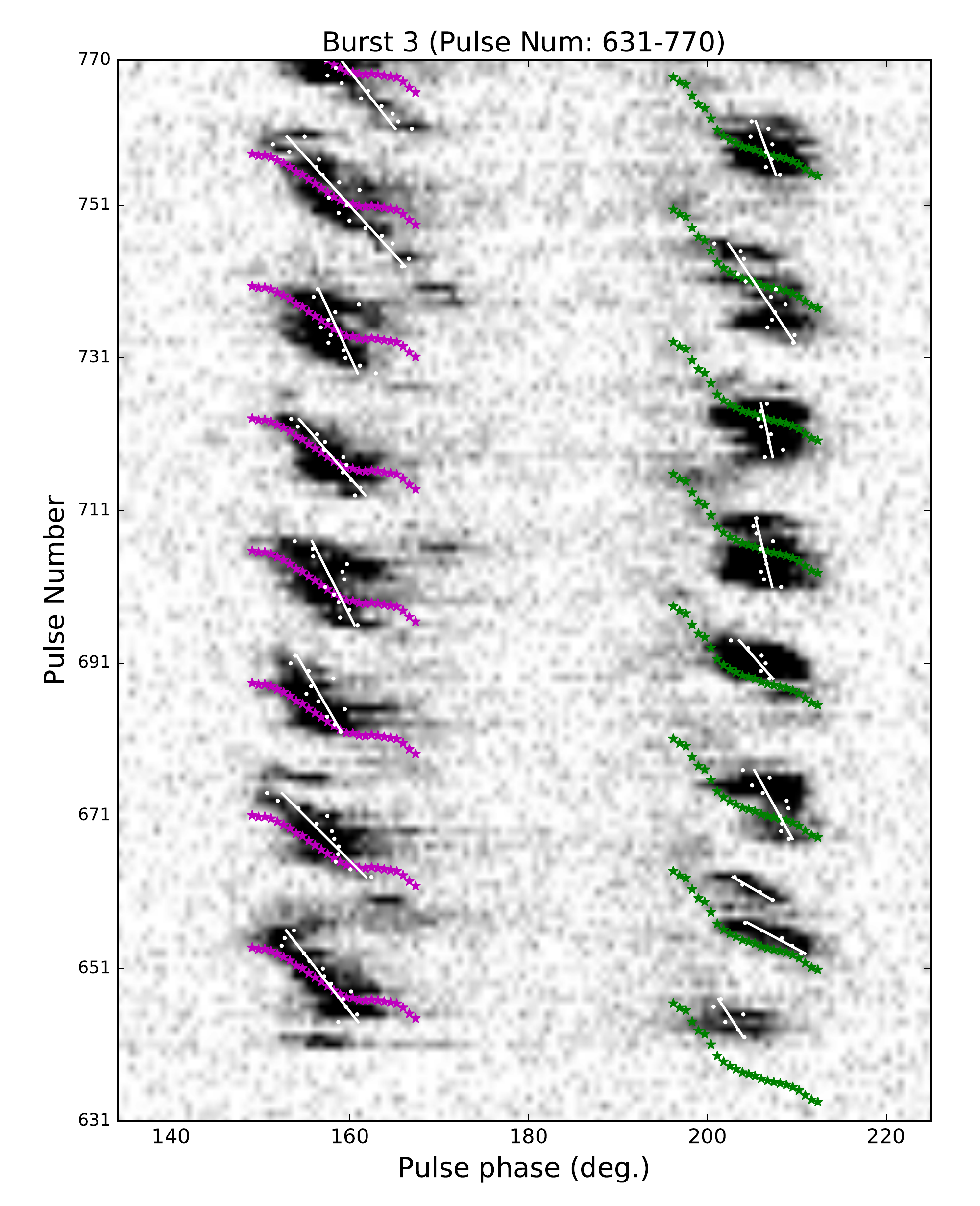} &
\includegraphics[width=8cm]{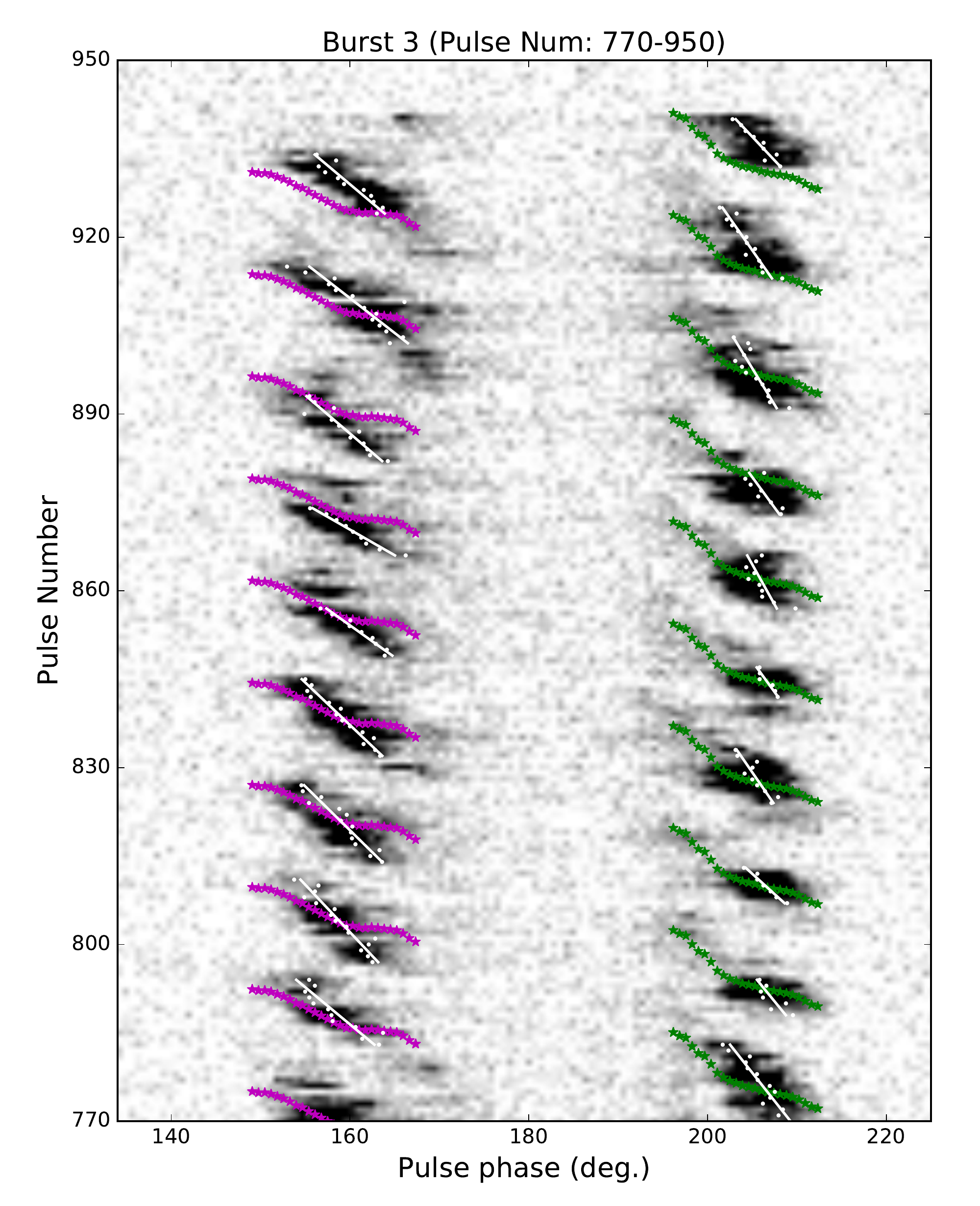} \\
\end{tabular}
\caption{Pulse stacks for the first half of Burst 3 (left panel) and the second half (right panel) with overlaid fits to the drift band structure. The white points are the centroid of gaussian profiles to the subpulses in each pulse and the white lines are a weighted least-squares fit of a straight line to the centroid points for a given drift band. The purple and green points result from fits of a cosine function over the whole burst to each pulse phase bin across the profile components assuming constant $P_3$ values for the leading and trailing components.  The derived cosine phases were converted to give the locus of the cosine peak near $t_0$ (pulse 770) and then replicated across the burst. See text for further details.}
\label{fig.B3_slopes}
\end{figure*}

Figure~\ref{fig.C12_6Burst} shows that the slopes of drift bands vary substantially from band to band within a burst, between bursts, and for the two main components, C1 and C4. To make this quantitative, we have fitted a single gaussian to the intensity of each subpulse group representing a given drift band in each pulse. We then do a weighted fit of a straight line to the centroid phase of the fitted gaussians for a given drift band to measure the drift rate or band slope $\Delta\phi = P_2/P_3$, and its uncertainty. Note that, with this definition, $\Delta\phi$ is zero for a vertical band in a stack plot. Histograms of the band slopes for components C1 and C4 for Burst 3 and for all other bursts combined are given in Figure~\ref{fig.BursHis}. The fitted centroid points and linear fits to these points are shown on Figure~\ref{fig.B3_slopes} for Burst 3.

These histograms confirm that observed band slopes or subpulse drift rates are quite variable, especially for the shorter bursts, and are systematically different for components C1 and C4. Positive drift rates are seen only in the short bursts, specifically Bursts 4, 5a and 6. 

\begin{figure}
\centering
\includegraphics[width=1\linewidth]{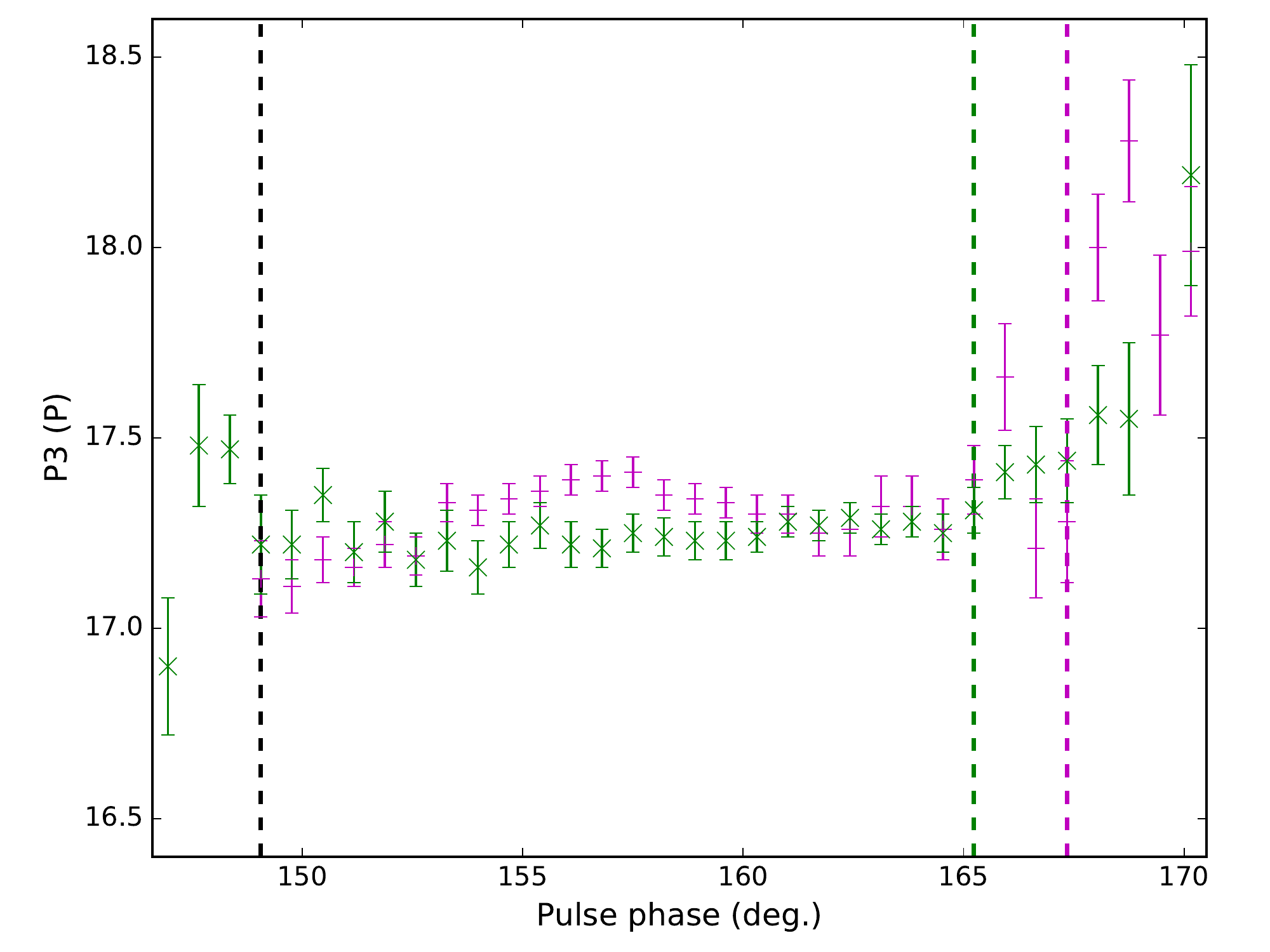}
\caption{Variation of $P_3$ across the leading components (C1 and C2, purple +) and the trailing components (C3 and C4, green $\times$) for the burst number 3. For the trailing components, the actual pulse phase is $47\degr$ more than that indicated. The vertical dashed lines mark the boundaries of the significant emission. The left boundary applies to both the leading and trailing components. }
\label{fig.B3_P3}
\end{figure}

To further investigate the characteristics of the drifting subpulses in PSR J1926$-$0652, we undertook a Fourier analysis of the longest burst, Burst 3, which also has the most regular drifting. The frequency (or equivalently $P_3$ modulation period), phase and amplitude of a cosine function were fitted to the pulse intensities across the burst for each pulse phase bin. Pulse 770, near the center of the burst and at the boundary of the two panels in Figure~\ref{fig.B3_slopes}, was adopted as the reference time, $t_0$, for the cosine fit. Figure~\ref{fig.B3_P3} shows the variations of $P_3$ across the leading and trailing components. The weighted mean values of $P_3$ (averaged between the vertical dashed lines in Figure~\ref{fig.B3_P3}) are $(17.35\pm 0.04)P$ and $(17.31\pm 0.03)P$ for the leading and trailing components, respectively. The difference between these values is of marginal significance and so we adopt a mean $P_3$ of $(17.33\pm 0.03)P$ for the whole profile. 

Using this mean $P_3$ value, we fit for the cosine phase at $t_0$ across the leading and trailing components. For the drift band closest to the reference time $t_0$, the time of the cosine maximum for a given pulse phase bin is  given by:
\begin{equation}
    t_{\rm max} = t_0 - \phi_0 P_3
\label{eq:mod_phs}
\end{equation}
where $\phi_0$ is phase of the modulation at $t_0$, and $t_0$ and $t_{\rm max}$ are expressed in units of pulse period.\footnote{Note that for a band drifting toward earlier pulse phases (as in this case) the Fourier ($P_3$) phase at $t_0$ is an increasing function of pulse phase. The minus sign in Equation~\ref{eq:mod_phs} then implies that the slope of the drift bands and hence $P_2$ are negative for this pulsar. This Fourier phase convention is opposite to that adopted by \citet{Weltevrede+16} although the convention on the sign of $P_2$ is the same.} 
Figure~\ref{fig.B3_slopes} shows the locus of the peak of the cosine function as a function of pulse phase for both the leading and trailing components.  The locus of the modulation peak was then replicated for all drift bands in the burst using the same mean value of $P_3$ for both the leading and trailing components.

The rate of Fourier phase drift is fairly stable through the main C1 and C4 components, about $-1.35 \degr/P$ and $-1.95\degr/P$ respectively. Given the mean $P_3$ value, these slopes correspond to $P_2 = -23\degr$ and $-34\degr$ respectively, with an uncertainty of about $1\degr$. However, the drift rate is quite non-linear across each component, appearing to flatten toward the component edges. Furthermore, the modulation phases of the inner Components 2 and 3 do not lie on the extrapolation of the phase variations in the main components. The modulation phases of the main components also differ, with Component 4 reaching its maximum amplitude about $90\degr$ in modulation phase (i.e., $0.25P_3$) later than Component 1. This means that for most pulses there is emission in one or both components although there are pulses with no significant emission. These are not ``nulls" in the usual sense, but just a consequence of the periodic modulations in the various components.  

It is clear that the band slopes derived from the Fourier analysis are very different to those derived from the direct gaussian fitting to the band profiles and illustrated in Figures~\ref{fig.BursHis} and \ref{fig.B3_slopes}. Since $P_3$ is relatively stable, this implies a similarly different distribution of the derived $P_2$ values. These results will be discussed further in Section~\ref{sec:discussion}. Estimates of $P_2$ and $P_3$ can also be obtained by computing fluctuation spectra. A description of such an analysis is provided in the Appendix~\ref{sec:2DFS}, showing that  consistent.

\begin{figure}
\centering
\includegraphics[width=1\linewidth]{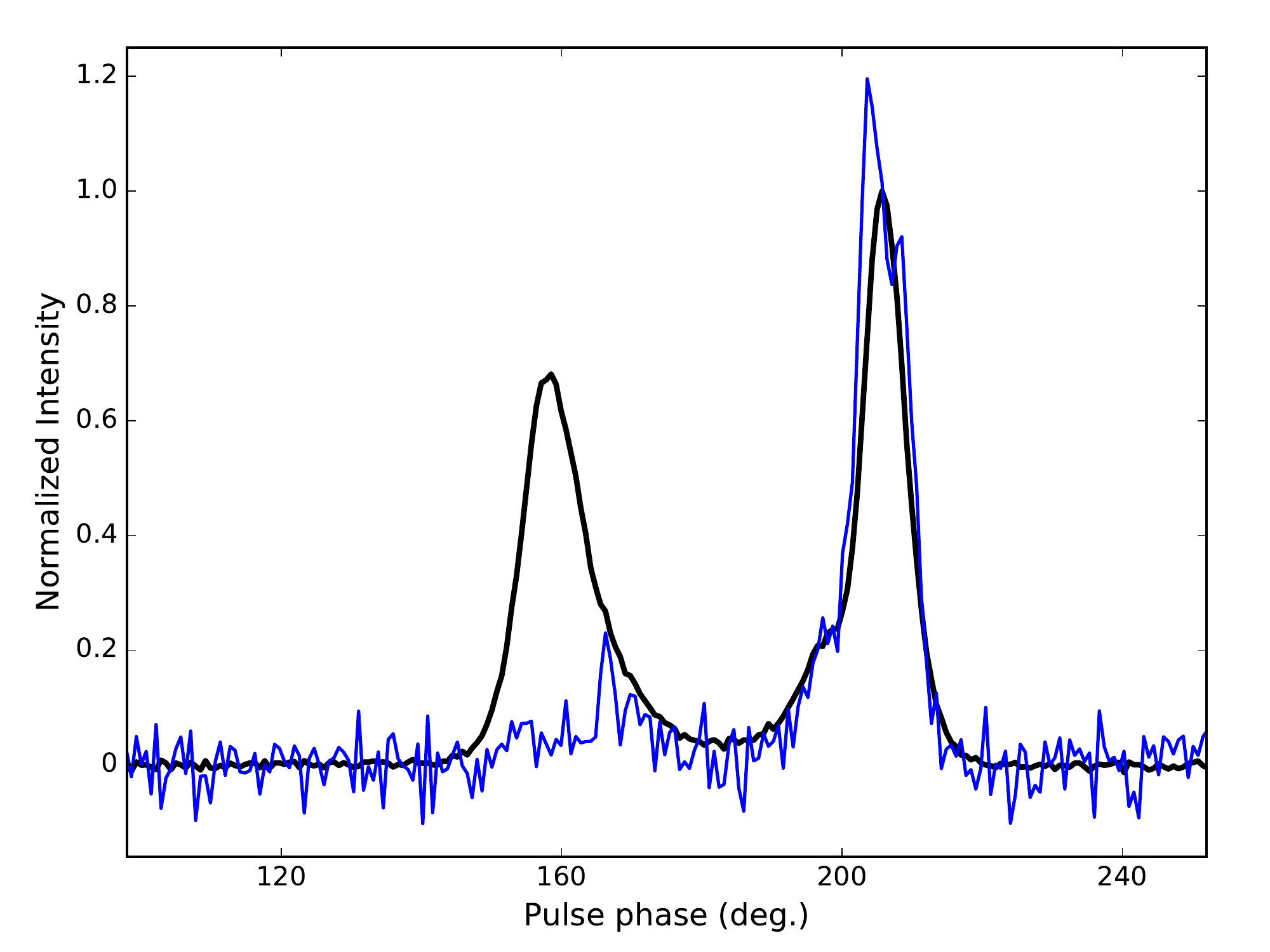}\\
\caption{Mean pulse profile averaged over all bursts (thicker black line) and the average profile for the last detectable pulse of each burst (thinner blue line). The mean burst profile peak is normalised to 1.0.  The profiles were averaged across the FAST observing band. }
\label{fig.lastpulse}
\end{figure}

Inspection of Figure~\ref{fig.C12_6Burst} indicates that the trailing pulse components (C3 and C4) are always detectable in the last pulse before a null event, whereas the leading pulse components generally are not.  In Figure~\ref{fig.lastpulse} we plot the mean profile of the last active pulse (LAP) of each burst (taking Bursts 5a and 5b separately) and the mean pulse profile over all bursts.  The LAP average profile is clearly dominated by the trailing components, although there is occasional emission for components 1 and 2, for example in Burst 3. Within the uncertainties, the LAP emission for C3 and C4 has the same shape as the mean profile over all burst emission and a similar amplitude. To quantify the significance of the shape change we have carried out 500,000 trials in which we have summed seven randomly-selected pulses (from the ``on" or burst states) to form an integrated profile.  The strength of the leading components relative to the trailing components was determined for each trial by calculating the area beneath the components using \textsc{psrsalsa}  \citep{Weltevrede+16}.  Out of the 500,000 trials, only one had a more extreme ratio than is observed in Figure~\ref{fig.lastpulse} (0.154), thereby confirming that the weakness of the leading components in the last active pulse prior to a null is not a chance result.\footnote{A slightly higher, but still very low, ratio is obtained by taking six randomly-selected pulses and treating Burst 5 as one burst. The deviation of the LAP profile from the average is still highly significant in this case.} The first detectable pulse of each burst is not systematically different from an average pulse, with two bursts starting with the leading component (e.g., Bursts 5a and 5b), one with the trailing component (Burst 4) and two with both components starting at the same time (Bursts 3 and 6).

\section{The long-term timing and emission properties}
\subsection{Timing solution}

\begin{figure}
\centering
\includegraphics[width=1\linewidth]{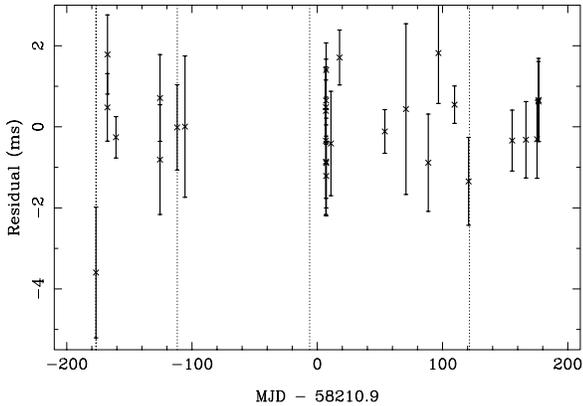}\\
\caption{The timing residuals corresponding to 353\,days of Parkes monitoring observations in the 20\,cm observing band. Note that the pulsar was only detected in 29 of the Parkes observations. It was not detected at the epochs where we have drawn a vertical, dotted line.  Three such observations were recorded on the same day (at $-$176 days on the Figure).}
\label{fig.TimingResidual}
\end{figure}

Timing residuals were formed using the long-term Parkes monitoring observations. We used the \textsc{tempo2} \citep{Hobbs+06} software package using the DE421 solar system ephemeris and the TT(TAI) time standard to obtain a phase-connected timing solution extending over 353\,days. The timing residuals are shown in Figure~\ref{fig.TimingResidual}. The nulling time-scale is too short to search for changes in the spin-down rate during such events and the pulse arrival times are modelled well using a very simple parameterisation of the pulsar.  The timing solution is presented in Table~\ref{tab.C12_parameters}. We also present parameters derived from the timing parameters, including the DM-based distance estimate from the \citet{Yao+17} model for the Galactic free-electron distribution, the pulsar's characteristic age ($\tau _{c} = P/2\dot P$) where $P$ is the pulse period and $\dot P$ is its first time derivative) and a representative surface-dipole magnetic field strength ($B_{s} = 3.2\times 10^{19}\sqrt{P\dot{P}}$\,Gauss) in the table. The mean flux density at 1400~MHz, $0.9\pm0.2$~mJy, was calculated by using the \textsc{psrchive} routine \textsc{psrflux} to give the flux density of each of the Parkes observations and then computing the mean and rms deviation of these values. The pulse widths are at 50\% of the peak amplitude and were computed from the mean profiles for the Parkes and FAST observations. 

We cross-correlated the pulse profile for each observation with our analytic template and inspected, by eye, the deviation from the scaled template and the observed profiles.  We found no evidence for pulse shape changes and therefore have no evidence that this pulsar exhibits discrete pulse-shape states.

\begin{table}[htp]
\begin{center}
\caption{Parameters for PSR~J1926$-$0652. Uncertainties in parentheses refer to the last quoted digit. All the parameters, apart from these indicated, were obtained from the Parkes observations.
\label{tab.C12_parameters}}
\begin{tabular}{ll}
\hline \\
 Timing model parameters:\\
 Right ascension (J2000) (h:m:s) &  19:26:37.11(3)\\
 Declination (J2000) (d:m:s)  &  $-$06:52:43.0(9)\\ 
 Galactic longitude ($\degr$) & 30.75\\
 Galactic latitude ($\degr$) &  $-$10.94\\
 Dispersion measure, DM (cm$^{-3}$~pc) &  84.7(9) \\
 Pulse period, $P$ (s) & 1.6088162910(6) \\
 Period derivative, $\dot P$ & $4.3(2)\times 10^{-16}$\\
 Epoch of period (MJD) & 58210 \\ 
 Time standard & TT(TAI)\\
 Time units & TCB \\
 Solar-system ephemeris & DE421\\
 RMS timing residual ($\mu$s) & 2830 \\
{\bf $\chi^{2}_{\rm red}$} & 0.98\\ \\
 \hline \\
 Derived parameters:\\
 Estimated distance$^{\rm a}$ (pc) & 5300\\
 Characteristic age (Myr) &  59.2\\
 Surface magnetic field strength (G) & $8.43\times10^{11}$\\ \\
 \hline \\
 Profile parameters: \\
 Mean flux density at 1400 MHz (mJy) & 0.9(2)\\
 50\% pulse width at 1400 MHz ($\degr$) & 45.0(4) \\
 50\% pulse width at 700 MHz$^{\rm b}$ ($\degr$) & 48.6(7)\\ 
 50\% pulse width at 350 MHz$^{\rm b}$ ($\degr$) & 56.1(7) \\ \\
 \hline \\
 Polarization parameters at 1400 MHz:\\
 Rotation measure (rad~m$^{-2}$) & $-$55(3)\\
 Linear polarization fraction ($L/I$) & 30\%\\
 Circular polarization fraction ($|V|/I$) & 1.6\%\\ \\
 \hline \\
 Long-term emission-state parameters:\\
 Longest ``on" duration (min.) & 20\\
 Mean ``on" duration (min.) & 5.9\\
 Standard Deviation ``on" duration (min.) & 3.6\\
 Longest ``off" duration (min.) & 93\\
 Mean ``off" duration (min.) & 20.3\\
 Standard Deviation ``off" duration (min.) & 20.6\\ \\
 \hline \\
 Observations:\\
 Number of observations & 35\\
 Date of first observation (MJD) & 58034\\ 
 Date of last observation (MJD) & 58387\\ 
 Total time span (days) & 353 \\ 
 \hline
\end{tabular}\\
$^{\rm a}$ Derived from the  \cite{Yao+17} model.\\
$^{\rm b}$ Derived from the FAST observation.\\
\end{center}
\end{table}

\subsection{Polarization properties and flux density}\label{sec:pol}

\begin{figure*}[thpb]
\centering
\begin{tabular}{ll}
\includegraphics[height=0.41\linewidth,width=8cm]{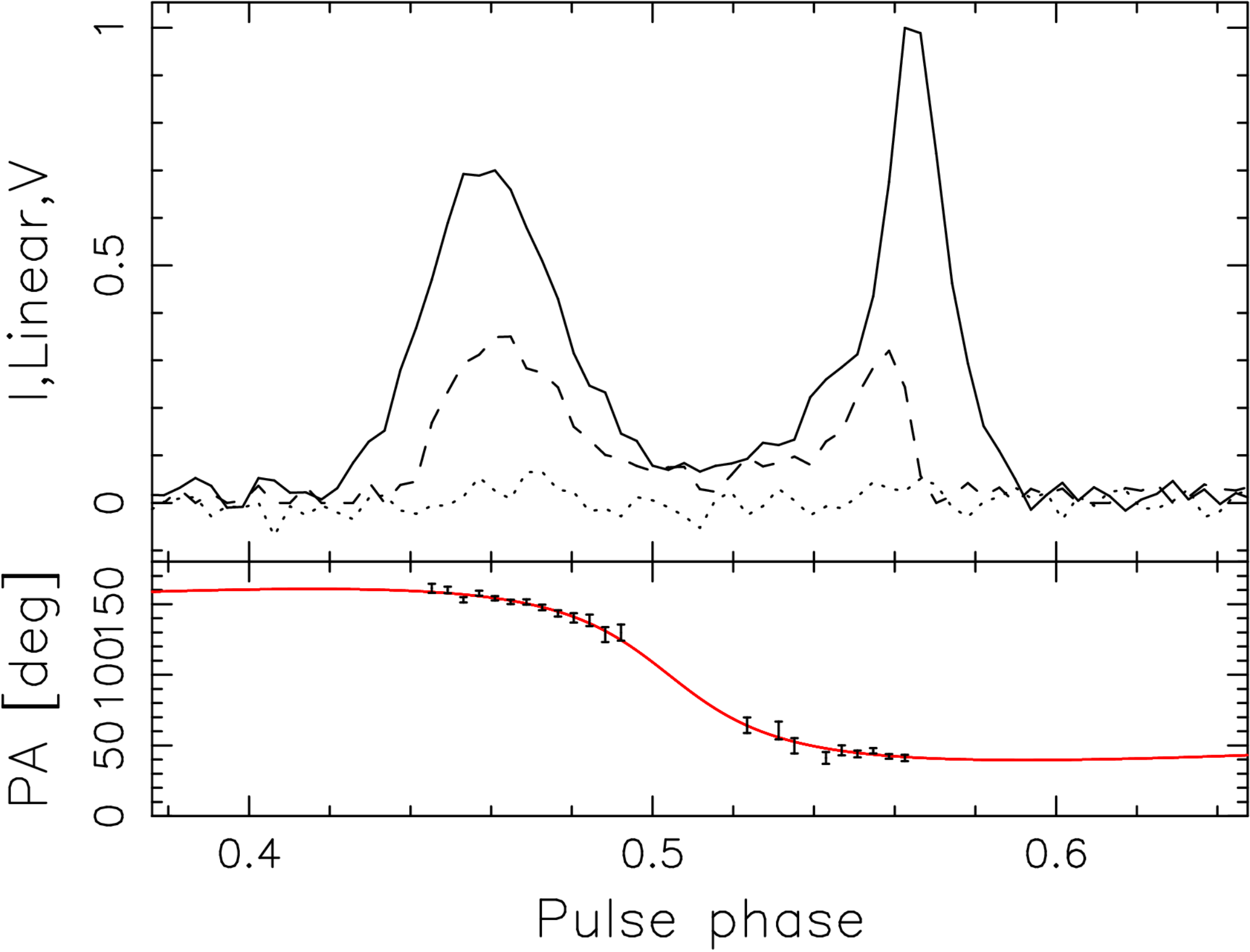} &
\includegraphics[height=0.41\linewidth,angle=0]{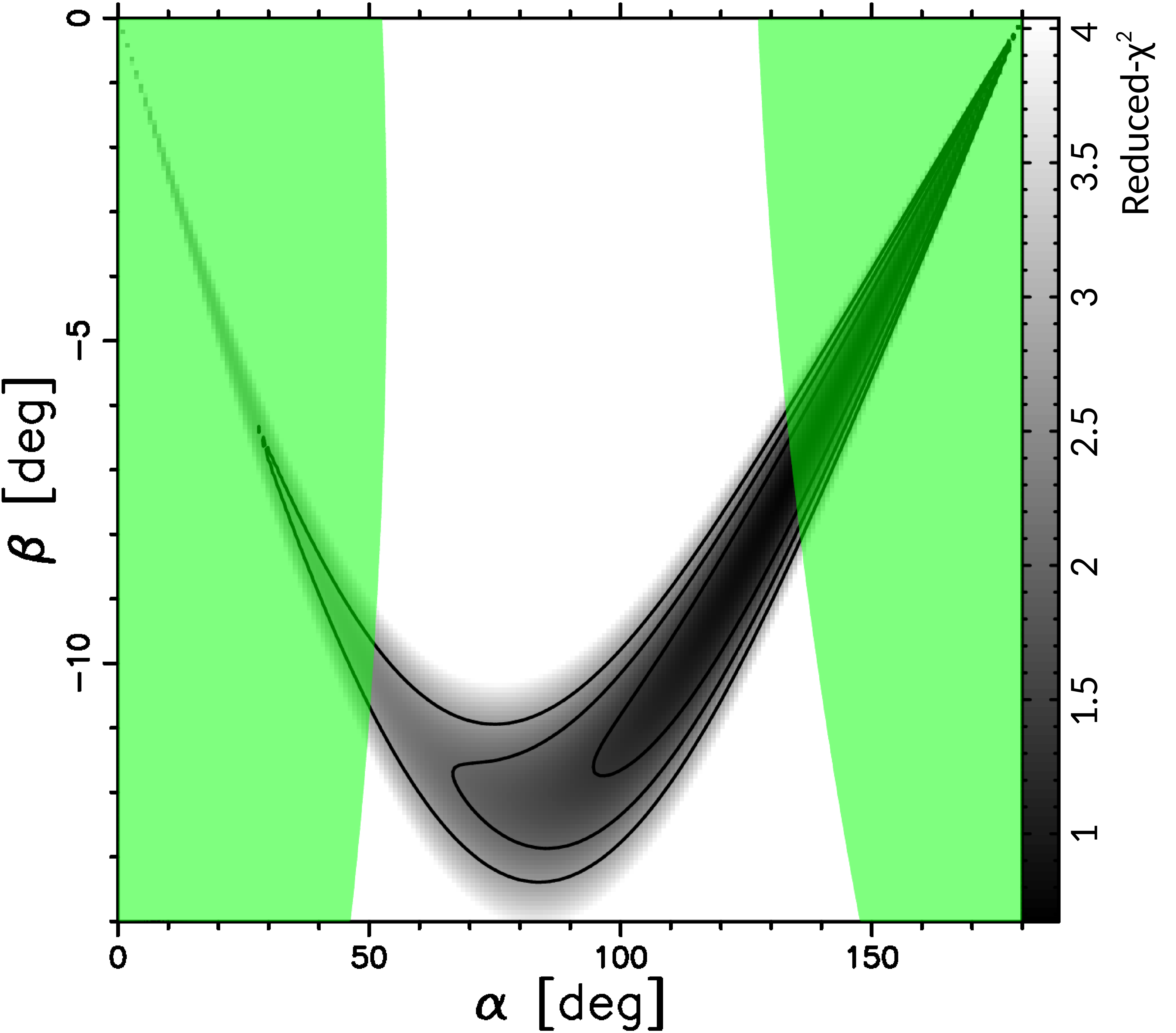} \\
\end{tabular}
\caption{\emph{Left:} Polarization profile at 20\,cm (1400~MHz). The black line is the mean flux profile, the dashed line is the linear polarization profile and the dotted line is the circle polarization profile in the top panel. The black dots, in the bottom panel, represent the linear polarization angles along with the best fit curve from the RVM fit shown as the red line. \emph{Right:} The results of fitting an RVM curve for each ($\alpha, \beta$) combination. The reduced chi-squared ($\chi^{2}$) of the fit is shown as the grey-scale, with the darkest value corresponding to the best fit. The black contour lines represent $1-\sigma$, $2-\sigma$, and $3-\sigma$ confidence boundaries.  The green regions show geometries allowed by the observed pulse width under certain assumptions -- see \S\ref{sec:geom}.}
\label{fig.C12_RVMfit}
\end{figure*}

To probe the polarimetric properties of the pulsar and to measure the average on-state flux density in the 20-cm observing band from the Parkes observations, we selected sub-integrations for observations in which emission was detected. Observations were aligned using the timing solution given in Table~\ref{tab.C12_parameters}, and then summed to produce a calibrated profile of the pulsar in the 20-cm observing band  using the \textsc{psrchive} software suite \citep{Hotan+04}. This summed profile was plotted using the \textsc{psrsalsa} software package \citep{Weltevrede+16} and is shown in the left panel of Figure~\ref{fig.C12_RVMfit}. We determined the rotation measure (RM) of the pulsar (${\rm RM} = -55\pm 3$\,rad\,m$^{-2}$) using the \textsc{rmfit} package. 

The average profile is moderately linearly polarized (dashed curve in the upper left panel of Figure~\ref{fig.C12_RVMfit}) with a fractional linear polarization of $30\pm1\%$. As is commonly observed in ``classic" double profiles \citep[e.g.,][]{lm88}, the degree of linear polarization is low at the profile edges and high in the bridge region. There is little evidence for significant circular polarization (dotted curve in the top panel of the left plot of Figure~\ref{fig.C12_RVMfit}). The  position angle (PA) curve of the linear polarization is shown in the bottom panel of left plot of Figure~\ref{fig.C12_RVMfit}. Its shape can be fitted using the rotating vector model (RVM; \citealp{Radhakrishnan+69}).  The fit is remarkably good, but the parameters are not well constrained\footnote{The PA curve in the Figure~\ref{fig.C12_RVMfit} uses values of magnetic inclination angle $\alpha = 158\degr$, impact parameter $\beta = -3\degr$, position-angle offset of $50\degr$ and fiducial-plane angle of $181\degr$}. In the right-hand panel of Figure~\ref{fig.C12_RVMfit} we show the reduced $\chi^2$ values of the fit as a function of $\alpha$ and $\beta$. The magnetic inclination angle, $\alpha$, is practically unconstrained and, from the RVM fit alone, we can only conclude that $\beta < 13^\circ$.   We describe more constraints on these parameters in the discussion section.

\subsection{Long-term, on-off time scale}

The Parkes observations, typically $\sim$1\,hr in duration (but sometimes as long as $\sim 7$\,hours) show that the time period during which the emission remains on or off lasts for tens of minutes. Parameterizing the exact on-off time scale is non-trivial as the emission state may have only switched once during a given observation (and so we have no prior information on how long it was on or off before or after the observation). Also some of the observations were affected by RFI, which was often so strong that we were unable to determine whether the emission switched states during the RFI. Our sub-integration time is 30\,s for the Parkes observations and so we assume that the emission remains on (or off) when RFI is affecting our data for less than four sub-integrations (2 minutes).  Similarly, calibration observations (lasting a couple of minutes) were carried out regularly through long observations of the pulsar and we assumed that the pulsar remained in a single state throughout those calibration observations. With these assumptions the maximum on-state duration is $\sim 20$ minutes. The maximum off-state duration is $\sim 93$ minutes. {The distribution of on and off state durations are quantified statistically in Figure~\ref{fig.on-off_his} and Table~\ref{tab.C12_parameters}\footnote{Without the assumption that the pulse stays on or off across small time gaps, we then obtain maximum on and off durations of 13.5 and 36\,min respectively}. 

\begin{figure}
\centering
\includegraphics[width=1\linewidth]{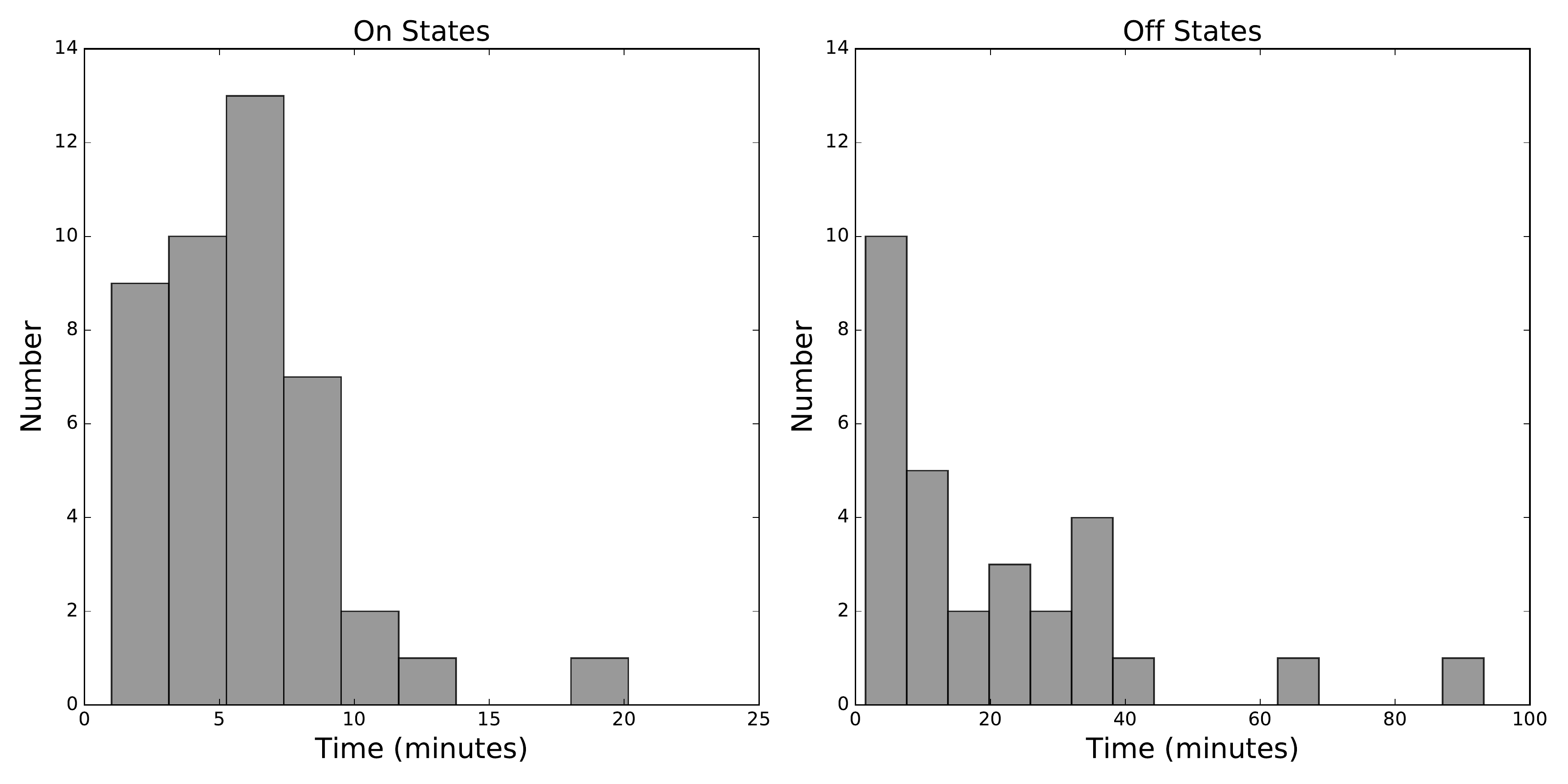}\\
\caption{Histograms of the observed on (left panel) and off (right panel) state durations.}
\label{fig.on-off_his}
\end{figure}

\section{Discussion}\label{sec:discussion}

\subsection{The discovery of PSR~J1926--0652}

PSR~J1926$-$0652 is a relatively bright pulsar and can be detected with the Parkes telescope within a few minutes.  We therefore wished to understand why this pulsar had not been discovered by previous surveys.  We checked the Parkes data archive \citep{Hobbs+11} and downloaded  previous search mode observations (that were not embargoed) that were observed at positions within 10\,arcmin of the known pulsar position. We identified two observations\footnote{ P309: An intermediate-latitude millisecond pulsar survey}, each of 4.3\,min duration, and searched for the pulsar using \textsc{ presto}\footnote{https://www.cv.nrao.edu/~sransom/presto/} \citep{Ransom+01}. We did not detect the pulsar, but this is not surprising as we know PSR~J1926$-$0652 has a nulling fraction of $\sim 75$\% and an average off-state duration of 20\,minutes.  The ``intermittency" (i.e., the on/off time-scale) in the pulse emission timescales is similar to that seen in other pulsars such as PSR~J1717$-$4054 \citep{Kerr+14, Young+15}. Clearly it is likely that there are many such pulsars remaining to be discovered by repeated observations of the same sky position. 

\subsection{The emission and viewing geometry}\label{sec:geom}

The pulse profile has two primary components (C1 and C4), two inner components (C2 and C3) and bridge emission.  Various generic models for pulse profile shapes have been suggested including a core and one or more cones, randomly distributed patches or patchy cones (see, e.g., \citealp{Karastergiou+07} for a review and an empirical model).  Our work does not explicitly confirm, nor rule out, any particular model, but we note that the profile component separations decrease as expected for higher-frequency emission occurring lower in the pulsar magnetosphere.  As shown in Section~\ref{sec:pol} the position angle of the linear polarization is also remarkably well fitted by a rotating vector model. However, the magnetic inclination angle, $\alpha$, is unconstrained.

Following the description in \citet{Rookyard+15}, further constraints on the viewing geometry can be obtained by making various assumptions.  The relatively large width of the profile (taken to be $60\degr\pm5\degr$ based on the Parkes data) implies either that $\alpha$ is small, or that the emission comes from high up in the magnetosphere.  This height can be constrained since the RVM inflection point occurs very close to the mid-point of the profile. \citet{Rookyard+15} considered how the inflection point can be delayed relative to the position of the fiducial plane.  The upper limit of this delay for our data is only $\sim 15^\circ$, which implies an emission height lower than 5000\,km.   In Figure~\ref{fig.C12_RVMfit} we have identified values of $\alpha$ and $\beta$ that can produce a pulse of the measured width. These are shown in the green areas and suggest that the magnetic axis is relatively aligned with the pulsar's rotation axis with $\alpha  < 55^\circ$.

\subsection{The subpulse drifting and nulling phenomena}

As Figures~\ref{fig.C12_SP+profile} and \ref{fig.C12_6Burst} clearly show, PSR J1926$-$0652 exhibits drifting subpulses. However their properties are complex. The drifting is more regular in the longer bursts and, specifically for the longest Bursts 1 and 3, is more regular in the leading component C1, compared to the trailing component C4.   Figure~\ref{fig.B3_slopes} shows that the modulation phases for the interior components C2 and C3 do not lie on the extrapolation of the band slopes for C1 and C4 respectively. The modulation phases shown in Figure~\ref{fig.B3_slopes} appear to smoothly join the inner C2, C3 components to the outer C1, C4 components, but this may be an artifact of the smoothing in time over the burst inherent in the Fourier analysis. There is no significant $P_3$ modulation for the bridge region between components C2 and C3. 

As mentioned in Section~\ref{sec:singlePulse}, the band slopes obtained from the gaussian fits to subpulse profiles and the Fourier analysis are systematically different, especially for the trailing component C4. This is most clearly illustrated in the longest burst, Burst 3 (Figure~\ref{fig.B3_slopes}). The Fourier band slopes tend to be flatter (larger absolute values) and, similarly, the derived $P_2$ values have larger absolute values. The reasons for this are not entirely obvious. The Fourier method averages over the whole burst, while the gaussian fits are independent for each drift band. There are significant variations in band spacing ($P_3$) from band to band, especially at the beginning and end of the burst, where the actual drift band times differ substantially from the Fourier phase predictions assuming a constant $P_3$. Additional band structures not described by the Fourier model exist, most notably the additional bands seen in C1 at pulse number 676 and in C4 at pulse number 661. 

On the other hand, there is a degree of subjectivity involved in choosing the subpulse structures to fit with the gaussian analysis. For example, it could be argued that there are independent double bands for C1 around pulse numbers 704 and 738. For C4, the drift structure is not so clear and the Fourier phases are evidently dominated by a few relatively flat bands, for example, around pulse numbers 655, 792 and 809. Both methods have their strengths but, unless the drift-band structure is very regular, they can give quite different results. 

The carousel model, originally proposed by \citet{{Ruderman+75}}, is widely used to interpret drifting subpulses. In this model, emission is produced from a series of ``sparks'' that circulate at a fixed period around the magnetic axis. As these sparks rotate past the observer's line of sight, they give rise to the characteristic drifting subpulses seen in many pulsars. 

PSR J1926$-$0652 has four profile components, each of which has distinct subpulse behaviour. Four components would naturally arise if there is emission from a second, inner carousel of sparks. Within the uncertainties, all components share the same periodicity ($P_3 \approx 17.3P$), suggesting that any such nested carousels are phase-locked in the sense that they have the same rotation period and the same number of sparks. Nested phase-locked carousels have been proposed previously, for example to explain the drifting subpulses of PSR~B0818$-$41 \citep{Bhattacharya+09}.  

However, for PSR J1926$-$0652 there are a number of features that do not fit naturally into such a carousel model. For example, there are significant variations in band spacing ($P_3$) between different drift bands for a given component. Also, there are clear extra drift bands that are not part of the regular $P_3$ modulation. 

Other models for drifting subpulses also exist. For instance, \citet{Gogoberidze+05} suggest the possibility that the drifting subpulses result from the modulation of radio emission by magnetospheric oscillations.   Such resonances that beat with the rotation of the pulsar may provide more natural explanations than carousel models for apparently complex phenomena such as harmonically related drift rates as seen in, e.g., PSRs~B0031$-$07 and B2016$+$28 \citep{Taylor+75}, variable and even reversing drift rates as seen in PSR~B0826$-$34 \citep{Gupta+04}  or opposite drift directions in different pulse components such as those observed in PSRs J0815+0939 \citep{clm+05} and B1839$-$04 \citep{Weltevrede+16}.

As described in Section~\ref{sec:singlePulse}, the first pulses observed after a nulling event are comparable to a typical on-state pulse. However, we have shown that the last active pulses prior to a nulling event are significantly different in that the leading component is significantly weaker than the trailing pulse components (Figure~\ref{fig.lastpulse}). The leading component could be weaker before a null as it fades away, but we see no evidence of such fading in our observation.  In contrast we see relatively strong emission in this component at the end of Burst 3. A second possibility is that the nulling events occur when the leading component is at (or near) its weakest point in the modulation cycle.  This is similar to that observed by \cite{Gajjar+17} for PSR~J1840$-$0840 which consistently enters the null state at the end of a drift-band in one of its profile components.  For PSR~J1926$-$0652 we can not make such a definitive statement as we do see clear emission in the leading component in the last active pulse for Burst 3. However, we will show below that the drift-rate seems to change near the end of  this burst. 

Our results add to the menagerie of interesting phenomena relating to nulling and subpulse drifting and show that there does not seem to be a single, simple connection between nulling and drifting. For instance, \citet{Deich+86} found, in PSR~B1944$+$17, that null events were preceded by a decay in pulse intensity of around 50\% over about three pulse periods.  They also showed that, like PSR~J1926$-$0652, the last active pulses were quantitatively different in shape and more variable than other pulses.  Similarly individual pulses from PSR~J1727$-$2739 show a decay in intensity before a null event \citep{Wen+16}.  This pulsar also has two primary components and, like PSR~J1926$-$0652, the intensity of the leading component is weaker than the trailing component prior to a null event. In contrast to the pulsar described in this paper, the pulses immediately after a null event in PSR~J1727$-$2739 were also significantly different from typical pulses.  

 There is currently no single physical model that can explain all of these phenomena. Further observations which hopefully would capture even longer burst events will be needed to obtain a deeper understanding of this unusual pulsar and drifting and nulling in general.

\section{Conclusions}
We report here a pulsar discovery, namely PSR~J1926$-$0652, from the FAST radio telescope. Largely through FAST single pulse studies, aided by follow-up timing observations made by the Parkes telescope,  PSR~J1926$-$0652 is found to exhibit a plethora of emission phenomena, including nulling and subpulse drifting. Our main findings include that\\
\begin{enumerate}[1.]
\item PSR~J1926$-$0652 has a relatively long period of about 1.6 s and a mean 1400\,MHz flux density at about 0.9\,mJy.

\item The pulse emission switches off (nulls) about 75\% of the time, on time scales between four to 450 pulse periods, and with an average off-state duration of about 20 minutes.

\item PSR~J1926$-$0652 has two primary components, two weaker inner components, and bridge emission. The separation between components decreases in the higher-frequency bands, consistent with that expected from radius-to-frequency mapping. 

\item The average profile at 1400\,MHz is moderately linearly polarized with a fractional linear polarization of about 30\%. The magnetic inclination angle, $\alpha$, is poorly constrained from the position-angle fit alone. Future multiple-band and polarized single-pulse observations promise much better constraints.

\item  PSR~J1926$-$0652 exhibits complex drifting  subpulse properties. Its four profile components, each of which has distinct behavior. Four components would naturally arise if there is emission from a second, inner carousel of sparks. However, PSR~J1926$-$0652  possesses a number of features that do not fit into such a carousel model. Significant variations in band spacing ($P_3$) between different drift bands were seen for any given component. There are clear extra drift bands that are not part of the regular $P_3$ modulation.
\end{enumerate}

FAST continues to discover pulsars \citep{Li+18}, including bright ones that were probably missed by previous searches because of their nulling properties. We thus expect this work  to be just the first of many in reporting new noteworthy pulsars. For PSR~J1926$-$0652, we have only scratched the surface in terms of analyzing its emission mechanism. We have further observations planned with FAST to obtain more single pulse data sets and Parkes for continued timing and monitoring, particularly with the new ultra-wideband receiver. We will be able to calibrate future FAST data sets and therefore will be able to obtain high S/N single pulses that provide a more detailed insight in the single pulse emission mechanism.

Having a declination close to zero, this pulsar can be observed by almost all of the major radio telescopes. PSR~J1926$-$0652 holds the potential to help provide a coherent picture for explaining complex nulling and subpulse drifting, that do not fit easily into the simple carousel model.

\acknowledgments
\begin{acknowledgements}
This work is supported by National Key R\&D Program of China No. 2017YFA0402600, State Key Development Program for Basic Research (2015CB857100), the National Natural Science Foundation of China (Grant No. U1731238, 11725313, 11690024, 11743002, 11873067, U1731218，11565010, 11603046, U1531246) and the Strategic Priority Research Program of the Chinese Academy of Sciences Grant No. XDB23000000.

WWZ is supported by the Chinese Academy of Science Pioneer Hundred Talents Program.
QJZ is supported by the Science and Technology Fund of Guizhou Province (Grant Nos.(2015)4015, (2016)-4008)).
KL acknowledges the financial support by the European Research Council for the ERC Synergy Grant BlackHoleCam under contract no. 610058, the FAST FELLOWSHIP from Special Funding for Advanced Users, budgeted and administrated by Center for Astronomical Mega-Science, Chinese Academy of Sciences (CAMS), and the MPG-CAS Joint Project "Low-Frequency Gravitational Wave Astronomy and Gravitational Physics in Space".

The Parkes radio telescope is part of the Australia Telescope National Facility which is funded by the Australian Government for operation as a National Facility managed by CSIRO. Pulsar research at Jodrell Bank Centre for Astrophysics and Jodrell Bank Observatory is supported by a consolidated grant from the UK Science and Technology Facilities Council (STFC).




\end{acknowledgements}


\appendix
\section{Data access}\label{sec:data_access}
The raw data from the FAST telescope used in the single-pulse study in this paper are owned by the National Astronomical Observatories, Chinese Academy of Sciences, and are available from the corresponding author upon reasonable request. The observations from the Parkes telescopes have been obtained using project codes PX500 and PX501. Conditional on data embargoes, these data are available on CSIRO's data archive\footnote{https://data.csiro.au/} \citep{Hobbs+11}. We note that observations of the pulsar currently in the archive were recorded under the source name PSR~J1926$-$0649 (instead of the correct name of PSR~J1926$-$0652). The raw PX500 data have an 18-month embargo period, whereas the PX501 data have a 10-yr embargo period.

We have produced a publicly-downloadable data collection available from the CSIRO's data archive containing our processed data files. This data collection contains (1) FAST single-pulse data for PSR~J1926$-$0652 in four different frequency bands and (2) Parkes timing data at 20\,cm, including  pulse arrival times, the  arrival time file, the timing model file, the timing template file and the calibrated and summed profiles. This data collection is available from CSIRO Data Access Portal \citep{zhang+18}.

\section{Analysis using PSRSALSA}\label{sec:psrsalsa}
The software tools used to conduct the $P_3-$fold and fluctuation spectra analysis here are part of the \textsc{psrsalsa} package \citep{Weltevrede+16}, and are freely available online\footnote{https://github.com/weltevrede/psrsalsa}.

\subsection{$P_3-$fold}\label{sec:P3fold}
\begin{figure}[thp]
\centering
\includegraphics[width=0.65\linewidth]{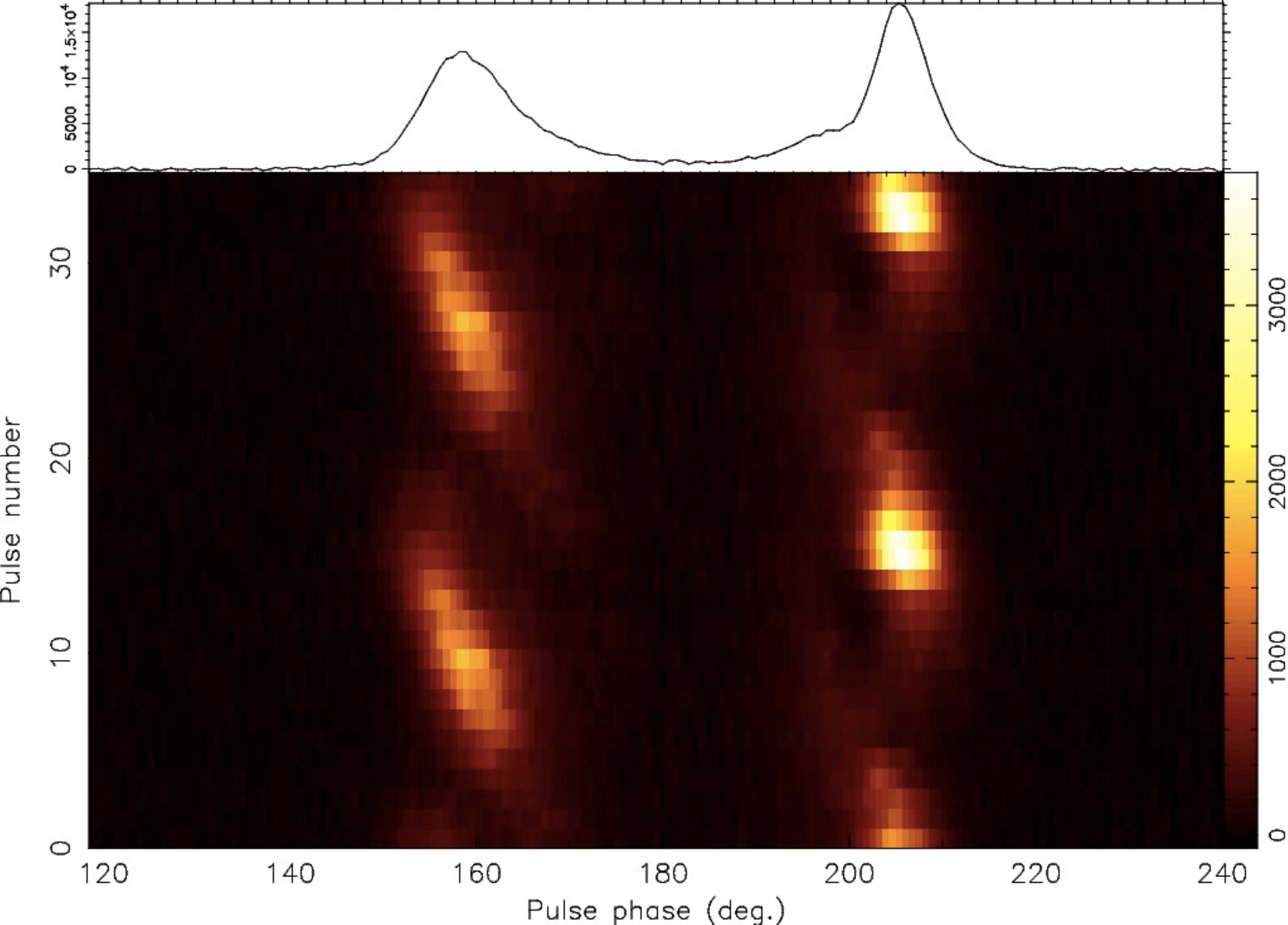}
\caption{A single $P_3-$fold for PSR~J1926$-$0652 covering the full 270$-$800\,MHz range for the longest observed burst sequence (Burst 3).}
\label{fig.C12_B3P3fold}
\end{figure} 

The single pulse data was folded at the identified period $P_3 = 17.33 P$, which was given in Section~\ref{sec:singlePulse}, using \textsc{psrsalsa} for Burst 3 (the longest observed burst sequence). This folding results in a high signal-to-noise representation of the average driftband, and permits more detailed studies of weak features in the drifting behaviour.

The $P_3$-fold (Figure~\ref{fig.C12_B3P3fold}) shows that leading component has an associated average driftband which is relatively steep, while the trailing component shows much shallower gradient. This is consistent with the measured $P_2$ value of trailing component being larger. In addition, the figure reveals that there two additional weak profile components with distinct drifting subpulse properties. The first of these two minor components (C2 as we described in the Section~\ref{sec:singlePulse}) can be associated with a small ``tail'' appearing around $170^{\circ}$ pulse longitude and pulse number $\sim18$. The second minor component (C3 as we described in the Section~\ref{sec:singlePulse}) around $195^{\circ}$ pulse longitude appears in the $P_3$-fold at pulse number $\sim23$.

\subsection{Fluctuation spectra analysis}\label{sec:2DFS}
The longitude-resolved fluctuation spectrum (LRFS) presents the spectral power of fluctuations as a function of rotational phase. The power in the LRFS can be used to quantify the longitude-resolved modulation index, which is shown, for Burst 3, as the points with error bars in the top-left panel of Figure~\ref{fig.C12_2DFS-B3}, together with the pulse profile.

A two-dimensional Fourier Transform of the pulse stack produced the two-dimensional fluctuation spectrum (2DFS) for Burst 3 in panel b of  Figure~\ref{fig.C12_2DFS-B3}. The 2DFS of the leading and trailing components are shown separately, for the whole band between 270 and 800 MHz. More examples of such statistical analyses can be found in, e.g., \citet{Edwards+03} and \citet{Weltevrede+06a, Weltevrede+06b}.

The vertical frequency axis of  both the LRFS and 2DFS corresponds to $P/P_{3}$, where \emph{P} denotes the rotational period of the pulsar. The LRFS shows a clear spectral feature at P/$P_{3}$ $\simeq$0.058 cycles per period ($cpp$) for both components. This spectral feature corresponds to the pattern repetition period of the drifting subpulses $P_{3}$ $\simeq 17P$ that can also be identified by eye in the pulse stack. In addition to this well-defined spectral feature, there are two weaker peaks for the leading component at $\simeq 0.067\,cpp$ corresponding to $P_{3} \simeq 15P$ and at $\simeq 0.043\,cpp$ corresponding to $P_{3} \simeq 23P$ (see in the top part of panel b) and one weaker peak for the trailing component at $\simeq 0.068\,cpp$ corresponding to $P_{3} \simeq 15P$ (see in the bottom part of panel b). These results indicate variations in the $P_{3}$ parameter for Burst 3. The horizontal axis of the 2DFS denotes the pattern repetition frequency along the pulse longitude axis, expressed as P/$P_{2}$. Following the description in \citet{Weltevrede+06b}, we have measured the $P_{2}$ and $P_{3}$ for the well-defined spectral feature for the two components. This gives  $P_{2} \simeq -29_{-2}^{+3}$\,deg and $P_{3} \simeq 17\pm 0.5\,P$ for the leading component  and  $P_{2} \simeq -42_{-10}^{+5}$\,deg and $P_{3} \simeq 17\pm 0.5\,P$ for the trailing component. We note that the quoted errors do not capture the fact that there is a high variability in the drift band shapes, and only a relatively small number of drift bands are observed. These results are consistent, but less accurate with the our Fourier analysis result ( $17.35\pm 0.04\,P$ and $17.31\pm 0.03\,P$ for the leading and trailing components respectively), which was given in Section~\ref{sec:singlePulse}.

\begin{figure}[thp]
\centering
\includegraphics[width=0.85\linewidth]{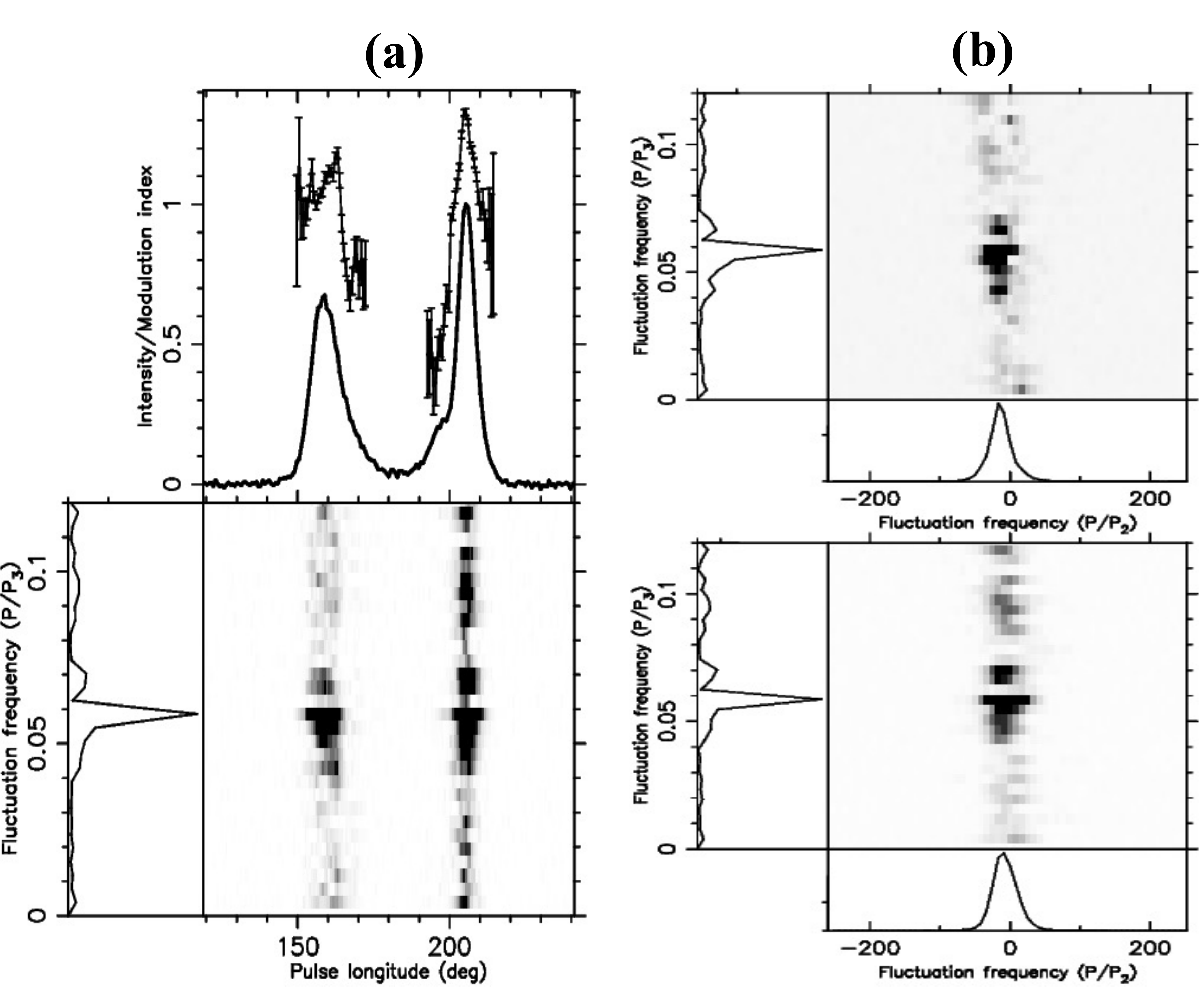}
\caption{Fluctuation analysis of the emission in the Burst 3 state. (a)The top panel shows the integrated pulse profile (solid line) and the longitude-resolved modulation index (solid line with error bars). Below this panel the LRFS is shown with on its horizontal axis the pulse longitude in degrees, which is also the scale for the abscissa of the plot above. (b) Analyse for each component: The top panel is the 2DFS of leading component and side panels showing the horizontally \emph{(left)} and vertically \emph{(bottom)}, integrated power. The bottom panel is the 2DFS of trailing component. Note: there are 300 pulses during Burst 3 (pulse number from 641 to 940). In order to make the most of pulses and give a high resolution, we used the last 256 successive pulses (pulse number from 641 to 896) in Burst 3 for our fluctuation analysis here. We also note that these fluctuation spectra show only part of the full spectra (which extend up to $P/P_3=0.5$ cpp).}
\label{fig.C12_2DFS-B3}
\end{figure}


\begin{thebibliography}{}

\bibitem[Asseo \& Melikidze(1998)]{Asseo+98}
Asseo E., Melikidze G. I., 1998, MNRAS, 301, 59

\bibitem[Baars et al.(1977)]{Baars+77}
Baasr J.W.M., et~al,1977, A\&A, 61, 99\\

\bibitem[Backer(1970b)]{Backer+70b}
Backer D.C., et~al, 1970, Nature, 228, 42\\

\bibitem[Bhattacharya et al.(2009)]{Bhattacharya+09} 
 Bhattacharya G., et~al, 2009, MNRAS, 398,1435\\


\bibitem[Biggs(1992)]{Biggs+92}
Biggs J. D, 1992, ApJ, 394, 574\\

\bibitem[Champion et al.(2005)]{clm+05} 
 Champion D., et~al, 2005, MNRAS, 363,929\\
 
\bibitem[Cordes(1978)]{Cordes+78}
Cordes J. M, 1978, ApJ, 222, 1006C\\

\bibitem[Cordes \& Shannon(2008)]{Cordes+08}
Cordes J. M., \& Shannon R. M., 2008, ApJ, 682, 1152\\

\bibitem[Deich et al.(1986)]{Deich+86}
Deich, W.T.S, et~al, 1986, ApJ, 300, 540\\

\bibitem[Deshpande \& Rankin(2001)]{Deshpande+01} Deshpande, A.A., \& Rankin. J.M., 2001, MNRAS, 322,438D\\

\bibitem[Drake \& Craft(1968)]{Drake+68}
Drake, F.D., \& Craft H.D., 1968, Nature, 220, 231\\

\bibitem[Edwards \& Stappers(2003)]{Edwards+03}
Edwards R. T., Stappers B. W., 2003, A\&A, 407, 273\\

\bibitem[Hobbs et al.(2006)]{Hobbs+06} 
 Hobbs G., et~al, 2006, MNRAS, 369, 655\\

\bibitem[Hobbs et al.(2011)]{Hobbs+11}
Hobbs G., et~al, 2011, Publ. Astron. Soc. Aust, 28, 202\\

\bibitem[Hotan et al.(2004)]{Hotan+04}
Hotan, van Straten \& Manchester, 2004, Publ. Astron. Soc. Aust, 21, 302\\

\bibitem[Gajjar et al.(2017)]{Gajjar+17}
Gajjar, V., et~al, 2017, ApJ, 850, 173\\

\bibitem[Gogoberidze et al.(2005)]{Gogoberidze+05} 
 Gogoberidze, G., et~al, 2005, MNRAS, 360, 669\\

\bibitem[Gupta et al.(2004)]{Gupta+04}
Gupta, Y., et~al. 2004,  A\&A, 426, 229\\

\bibitem[Karastergiou \& Johnston(2007)]{Karastergiou+07}
Karastergiou A \& Johnston S., 2007, MNRAS, 380, 1678\\

\bibitem[Kerr et al.(2014)]{Kerr+14}
Kerr C., et~al, 2014, MNRAS, 445, 320\\

\bibitem[Kramer et al.(2006)]{Kramer+06}
Kramer M., et~al, 2006, Science, 312, 549\\

\bibitem[Leeuwen et al.(2003)]{Leeuwen+03}
Van Leeuwen, et~al. 2003,  A\&A, 399, 223\\

\bibitem[Lewandowski et al.(2004)]{Lewandowski+04}
Lewandowski, W., Wolszczan, A., Feiler, G., Konacki, M., \& Sołtysin ́ski, T. 2004, ApJ, 600, 905\\

\bibitem[Li et al.(2018)]{Li+18}
Li D., et~al, 2018, IMMag, 19, 112L\\

\bibitem[Lyne \& Manchester (1988)]{lm88}
Lyne, A. \& Manchester, R., 1988, MNRAS, 234, 477\\

\bibitem[Lyne et al.(2010)]{Lyne+10}
Lyne, A., et~al, 2010, Science, 329, 408L\\

\bibitem[Manchester et al.(2013)]{Manchester+13}
Manchester D., et~al, 2013, Publ. Astron. Soc. Aust, 30, 17\\

\bibitem[McLaughlin et al.(2006)]{McLaughlin+06}
McLaughlin, M.A., et~al, 2006, Nature, 439, 817\\

\bibitem[Mitra(2017)]{Mitra+17}
Mitra D., 2017, JA\&A, 38, 52

\bibitem[Mitra et al.(2009)]{Mitra+09}
Mitra D., Gil J., Melikidze G., 2009, ApJ, 696, L141

\bibitem[Noutsos et al.(2008)]{Noutsos+08}
Noutsos A., et~al, 2008, MNRAS, 386, 1881\\

\bibitem[Palfreyman et al.(2018)]{Palfreyman+18}
Palfreyman J., et~al, 2018, Nature, 556,219

\bibitem[Qian et al.(2019)]{Qian+19}
Qian L, et~al, 2019, Sci. China-Phys. Mech. Astron, accepted.\\

\bibitem[Qiao et al.(2004)]{Qiao+04}
Qiao G. J, et~al, 2004, ApJ, 616L, 127Q\\

\bibitem[Radhakrishnan \& Cooke(1969)]{Radhakrishnan+69}
Radhakrishnan, V.; Cooke, D. J., 1969, ApJ, 3, 225R\\

\bibitem[Rankin (1986)]{Rankin+86}
Rankin, Joanna M., 1986, ApJ, 301, 901\\

\bibitem[Rankin et al.(2008)]{Rankin+08}
Rankin, Joanna M.; Wright, Geoffrey A. E., 2008, MNRAS, 385, 1923\\

\bibitem[Ransom(2001)]{Ransom+01}
Ransom, S. M. 2001, PhD, thesis, Harvard University\\


\bibitem[Rookyard et al.(2015)]{Rookyard+15}
Rookyard, S. C.; Weltevrede, P.; Johnston, S., 2015, MNRAS, 446, 3356R

\bibitem[Ruderman \& Sutherland(1975)]{Ruderman+75}
Ruderman M. A., Sutherland P. G., 1975, ApJ, 196, 51 (RS75)\\


\bibitem[Staveley-Smith et al.(1996)]{SS+96}
Staveley-Smith L., et~al, 1996, Publ. Astron. Soc. Aust, 13, 243\\

\bibitem[Sturrock(1971)]{Sturrock+71}
Sturrock P. A., 1971, ApJ, 164, 529\\

\bibitem[Taylor et al.(1975)]{Taylor+75}
Taylor, J.H., Manchester, R.N. \& Huguenin, G.R., 1975, ApJ, 195, 513\\

\bibitem[van Straten \& Bailes.(2011)]{Straten+11}
van Straten, W. \& Bailes, M., 2011, Publ. Astron. Soc. Aust, 28, 1\\

\bibitem[Wang et al.(2007)]{Wang+07}
Wang, N., Manchester, R. N., \& Johnston, S. 2007, MNRAS, 377, 1383\\

\bibitem[Wen et al.(2016)]{Wen+16}
Wen Z. G., et~al. 2016,  A\&A, 592, A127\\

\bibitem[Weltevrede et al.(2006a)]{Weltevrede+06a}
Weltevrede P., Edwards R. T., and Stappers B. W., 2006a, A\&A, 445, 243-272\\

\bibitem[Weltevrede et al.(2006b)]{Weltevrede+06b}
Weltevrede P., Wright G. A. E., Stappers B. W., and  Rankin J. M., 2006b, A\&A, 458, 269-283\\

\bibitem[Weltevrede(2016)]{Weltevrede+16}
Weltevrede P., 2016, A\&A, 590, A109\\

\bibitem[Wolszczan et al.(1981)]{Wolszczan+81} 
 Wolszczan A., Bartel N., \& Sieber W., 1981, MNRAS, 196, 473\\

\bibitem[Xie et al.(2019)]{Xie+19} 
 Xie Y. W., et al., 2019, preprint (arXiv:1903.01077)\\

\bibitem[Yao et al.(2017)]{Yao+17} 
 Yao J.M., Manchester, R. N., \& Wang N., et~al, 2017, MNRAS, 468, 3289\\
 
\bibitem[Young et al.(2015)]{Young+15} 
Young, N. J., Weltevrede, P., Stappers, B. W., Lyne, A. G., Kramer, M., 2015, MNRAS, 449, 1495\\

\bibitem[Zhang et al.(2018)]{zhang+18}
Zhang, L., et~al. 2018, Data files from Parkes and FAST for PSR J1926-0652. v1. CSIRO. Data Collection. https://doi.org/10.25919/5c354f21160e2\\

\bibitem[Zhu et al.(2014)]{Zhu+14}
Zhu,~W.~W., et~al. 2014,
ApJ, vol.~781, pp.~117\\

\end{thebibliography}
\end{document}